\documentclass[fleqn,usenatbib]{mnras}
\usepackage{newtxtext,newtxmath}
\usepackage[T1]{fontenc}
\usepackage{multirow}

\usepackage{graphicx}
\usepackage{amsmath}
\DeclareMathOperator*{\argmin}{argmin}
\DeclareMathOperator*{\argmax}{argmax}

\DeclareRobustCommand{\VAN}[3]{#2}
\let\VANthebibliography\thebibliography
\def\thebibliography{\DeclareRobustCommand{\VAN}[3]{##3}\VANthebibliography}

\newcommand{\msun}{M$_\odot$}

\usepackage[figuresright]{rotating}
\usepackage[normalem]{ulem}
\usepackage{xcolor}

\title[Core Evolution]{Turbulence, Coherence and Collapse: Three Phases for Core Evolution}

\author[S. S. R. Offner et al.]{
Stella S. R. Offner,$^{1}$\thanks{E-mail: soffner@astro.as.utexas.edu}
Josh Taylor,$^{1}$
Carleen Markey,$^{2}$
Hope How-Huan Chen,$^{1}$
Jaime E. Pineda,$^{3}$ \vspace{0.1in} \\
{\Large \normalfont Alyssa A. Goodman,$^{4}$
Andreas Burkert,$^{5}$
Adam Ginsburg,$^{6}$
Spandan Choudhury$^{3}$}
\\
$^{1}$Department of Astronomy, The University of Texas, Austin, TX 78712, USA\\
$^{2}$Department of Physics, Carnegie Mellon University, Pittsburgh, PA 15253, USA\\
$^{3}$Max-Planck-Institut f\"ur extraterrestrische Physik, Giesenbachstrasse 1, D-85748 Garching, Germany\\
$^{4}$Harvard-Smithsonian Center for Astrophysics, 60 Garden St., Cambridge, MA 02138, USA \\
$^{5}$University Observatory Munich (USM), Scheinerstrasse 1, 81679 Munich, Germany\\
$^{6}$Department of Astronomy, University of Florida, PO Box 112055, USA
}

\date{Accepted XXX. Received YYY; in original form ZZZ}

\pubyear{2022}

\begin{document}
\label{firstpage}
\pagerange{\pageref{firstpage}--\pageref{lastpage}}
\maketitle

\begin{abstract}
We study the formation, evolution and collapse of dense cores by tracking structures in a magnetohydrodynamic simulation of a star-forming cloud.  We identify cores using the dendrogram algorithm and utilize machine learning techniques, including Neural Gas prototype learning and Fuzzy $c$-means clustering, to analyze the density and velocity dispersion profiles of cores together with six bulk properties.  We produce a 2-d visualization using a Uniform Manifold Approximation and Projection (UMAP), which facilitates the connection between physical properties and three partially-overlapping phases: i) unbound turbulent structures (Phase I), ii) coherent cores that have low turbulence (Phase II), and iii) bound cores, many of which become protostellar (Phase III). Within Phase II we identify a population of long-lived coherent cores that reach a quasi-equilibrium state. Most prestellar cores form in Phase II and become protostellar after evolving into Phase III. Due to the turbulent cloud environment, the initial core properties do not uniquely predict the eventual evolution, i.e., core evolution is stochastic, and cores follow no one evolutionary path. The phase lifetimes are 
1.0$\pm$0.1$\times$10$^5$ yr, 
1.3$\pm$0.2$\times$10$^5$ yr, and 
1.8$\pm$0.3$\times$10$^5$ yr
for Phase I, II, and III, respectively. We compare our results to NH$_3$ observations of dense cores. Known coherent cores predominantly map into Phase II, while most turbulent  pressure-confined cores map to Phase I or III. We predict that a significant fraction of observed starless cores have unresolved coherent regions and that $\gtrsim 20$\% of observed starless cores will not form stars. Measurements of core radial profiles, in addition to the usual bulk properties, will enable more accurate predictions of core evolution.
\end{abstract}


\section{Introduction}
\label{sec:intro}

Since the first identification of dense cores in molecular line observations made by \citet{Myers_1983a}, astronomers have used the term {\it core} to describe the small \citep[$\sim$0.1 pc;][]{Jijina_1999}, roundish \citep[aspect ratio $\leq$ 2;][]{Myers_1991} and quiescent \citep[velocity dispersion nearly thermal;][]{Fuller_1992} blobs of gas that are likely progenitors of low-mass stars \citep{Pineda2022}.  Later observations further characterized most star-forming cores as gravitationally bound, if not collapsing \citep{Caselli_2002, Enoch_2008, Seo_2015}.  On the other hand, \citet{Shu_1987} formulated analytical star formation models and proposed an evolutionary sequence that describes the formation of protostars within cores through continuous accretion initiated by gravitational collapse and regulated by thermal pressure.  Efforts using both observations and numerical simulations to understand the evolution of dense cores have since been largely focused on how dense cores evolve from the point of time when they become self-gravitating ({\it prestellar cores}) to when protostars form within them \citep[{\it protostellar cores};][]{Li_2004, Tafalla_2004, McKee_2007, Offner_2008, Lada_2008, Kauffmann_2008, Rosolowsky_2008a, Dib_2010, Heigl_2016, CChen_2018,Grudic_2022}.

\citet{Barranco_1998} used observations of NH$_3$ hyperfine line emission to show that the line widths in the interiors of some dense cores are roughly constant at a value slightly higher than a purely thermal line width.  \citet{Goodman_1998} made observations of OH and C$^{18}$O line emission of dense cores and proposed that a characteristic radius exists where the scaling law between the line width and the core size changes from a power law to a virtually constant relationship.  \citet{Goodman_1998} found this characteristic radius to be $\sim$0.1 pc and called this change in the line width--size relation the transition to coherence.  A {\it coherent core,} defined by the transition to coherence, is hypothesized to provide the ideal low-turbulence environment for further star formation through gravitational collapse \citep{Goodman_1998, Caselli_2002}.  At around the same time, by measuring the near-infrared extinction, \citet{Alves_2001} found that the internal density structures of the dark cloud Barnard 68 are well described by a pressure-confined, self-gravitating isothermal sphere that is critically stable according to the Bonnor-Ebert criteria \citep{Ebert_1955, Bonnor_1956}.  Later observations of C$^{18}$O molecular line emission confirmed that Barnard 68 is a thermally supported dense core \citep[although a later study found evidence that Barnard 68 is possibly merging with a smaller structure, which would lead to destabilization and collapse;][]{Lada_2003b, Burkert_2009}.  Both the observation of coherent cores and the identification of a thermally supported dense core resembling a critical Bonnor-Ebert sphere provide important hints about the initial condition of dense cores before the formation of protostars within them.

Recent observational works have revealed that coherent cores are common in nearby molecular clouds.  \citet{Pineda_2010} made the first direct observation of a coherent core in the B5 region in Perseus.  \citet{Pineda_2010} observed NH$_3$ hyperfine line emission using the Green Bank Telescope (GBT) and resolved the transition to coherence across the boundary of the core.  Using Very Large Array (VLA) observations of the interior of the coherent core in B5, \citet{Pineda_2015} found substructures within the B5 coherent core that will likely form protostars in a freefall time of $\sim$40,000 yr.  \citet{Chen_2019a} identified a population of at least 18 coherent structures\footnote{In this work, coherent cores and coherent structures are used interchangeably to refer to dense cores defined by a transition to coherence.  The non-self-gravitating and pressure confined population of {\it droplets} identified by \citet{Chen_2019a} is a subset of coherent cores by this definition.  This slightly differs from the convention adopted by \citet{Chen_2019a}, where the term {\it coherent cores} specifically means self-gravitating coherent cores.  See \S3 in \citet{Chen_2019a}.} in Ophiuchus and Taurus using data from the GBT Ammonia Survey \citep[GAS;][]{GAS_DR1}.  These include droplets, a population of coherent cores that are not bound by self-gravity but are predominantly confined by the pressure provided by the turbulent motions of the ambient gas \citep{Chen_2019a}.  The non-self-gravitating droplets have density structures shallower than a critical Bonnor-Ebert sphere \citep{Chen_2019a} and sometimes show signs of internal velocity gradients that are likely the result of a combination of turbulent and rotational motions \citep{Chen_2019b}.  It was conjectured that these coherent structures, not bound by self-gravity, are either i) at an early stage of core formation, ii) an extension of the more massive coherent core population, or iii) transient.  Together, \citet{Pineda_2010} and \citet{Chen_2019a} revealed an entire population of coherent cores, ranging from self-gravitating and sometimes star-forming ones, including the B5 coherent core, to non-self-gravitating and predominantly pressure-confined droplets.  If coherent cores do indeed provide the necessary low-turbulence environment for star formation as hypothesized by \citet{Goodman_1998}, then an important question concerns whether there is an evolutionary relation between different flavors of coherent cores and between coherent cores and the better known pre-/protostellar cores.  Unfortunately, no coherent cores defined by a transition to coherence have been identified in simulations to date, although cores with subsonic velocity dispersions have been identified in simulations \citep[e.g.,][]{Klessen_2005,Offner_2008}.

In this work, we develop a method to identify, track and characterize the evolution of dynamic gas structures in simulations, which may be applied to other numerical models of star formation.  We aim to provide a complete picture of core formation and evolution that links turbulent molecular clouds to star-forming cores.  In particular, we aim to answer the following questions: i) how do cores form in a turbulent environment, ii) what role do coherent cores play in the star formation process, and iii) is there an evolutionary connection between coherent cores and pre-/protostellar cores?  To answer these questions, we carry out a comprehensive analysis of density structures in a magnetohydrodynamic (MHD) simulation of a turbulent molecular cloud.  We examine these structures as they evolve and move across the simulation without any prior assumptions regarding their internal structures.  We achieve this by utilizing unsupervised machine learning techniques, including Neural Gas prototype learning and Fuzzy $c$-means clustering.
  We then compare our results to cores identified in NH$_3$ in the Orion, Perseus, Taurus, Ophiuchus and Cepheus star-forming regions \citep{Kirk_2017a,Kerr_2019,Chen_2019a,Keown_2019}, including the known sample of coherent cores.

In \S\ref{sec:data}, we describe the MHD simulation and the set of observations that we compare to.  We then introduce our method to identify and track density structures in \S\ref{sec:analysis_id} and describe how we calculate core properties in \S\ref{sec:analysis_datadescription}. In \S\ref{sec:analysis_methodoverview} we present our approach to cluster cores using prototype learning and then describe the Uniform Manifold Approximation and Projection (UMAP) approach to visualize the result in \S\ref{sec:umap}.  We examine the properties of the core clusters ({\it phases}), investigate core evolution and compare to observations in \S\ref{sec:results}.  We discuss the implication of the phases for an evolutionary sequence in \S\ref{sec:discussion_evolution} and compare with star formation models in \S\ref{sec:discussion_formation}-\S\ref{sec:discussion_formationhigh}.  We discuss the implications for core observations in \S\ref{sec:discussion_observation} and caveats to our approach in \S\ref{sec:caveats}. We summarize our work in \S\ref{sec:conclusions}.

\section{Data}
\label{sec:data}

\subsection{Magnetohydrodynamic Simulation of Star Formation}
\label{sec:data_simulation}
We analyze the magnetohydrodynamic (MHD) simulation of a turbulent star-forming cloud previously presented in \citet{Smullen_2020}.  The simulation models a box of 5 pc on a side with periodic boundary conditions.  We focus on the data in the basegrid and first adaptive mesh refinement (AMR) level, which corresponds to a voxel size of $\sim$0.004 pc and is consistent with a Nyquist sampling of the beam size of observations used by \citet{Chen_2019a}.  The initial conditions of this simulation are identical to those of run W2T2 in \citet{Offner_2015}, where these conditions are chosen to model a typical nearby molecular cloud like the Perseus molecular cloud.  The simulation is run using the ORION2 code and includes ideal MHD, self-gravity and Lagrangian accreting sink particles \citep{Krumholz_2004,Li_2012,Orion_2021}.  The mean gas density of the simulation is $\rho_0 = 2.04\times10^{-21}$ g cm$^{-3}$, or $n\sim$430 cm$^{-3}$, where $n$ is the molecular hydrogen number density assuming a mean molecular weight per H$_2$ molecule of 2.8 a.m.u. \citep{Kauffmann_2008}.  The simulation begins with a uniform density, a uniform temperature of 10 K and a uniform magnetic field in the $z$-direction, $B_z$ = 13.5 $\mu$G.  The gas is then perturbed for two Mach crossing times by a random velocity distribution  with dispersion $\sigma_{3d}= 2.0\,$km\,s$^{-1}$  that corresponds to a flat power spectrum in Fourier space with $1 \leq k L / 2 \pi \leq 2$, where $k$ is the wavenumber and $L$ is the domain size.   At the end of the driving phase, the  gas reaches a turbulent steady state with a turbulent power spectrum $P(k)$ $\propto$ $k^{-2}$, plasma parameter (ratio of thermal pressure to magnetic pressure) $\beta$ = $8\pi \rho_0 c_s^2/B_z^2$ = 0.02, and virial parameter  $\alpha_{\rm vir} = 5 \sigma_{\rm 1d}^2 L /(2 G M_{\rm cloud}) = 1.0$ , where $c_s$ is the sonic speed and $M_{\rm cloud} \simeq 3800$\,M$_{\odot}$.  See \citet{Smullen_2020} for details.   We follow the cloud evolution for $6 \times 10^5$ yr and use simulation snapshots with time spacing $\Delta t\sim$1.5$\times$10$^4$\,yr for the analysis.

\subsection{Source Catalogs}
\label{sec:data_catalogs}
We compare the cores identified in the MHD simulation to cores observed using the NH$_3$ emission from the GBT Ammonia Survey \citep[GAS,][]{GAS_DR1}. These data were combined with different ancillary datasets to identify cores and derive their properties in several different star-forming regions. Note that each of the studies adopts a slightly different approach to core identification as we describe below. 

\subsubsection{Coherent Cores}

 \citet{Chen_2019a} identified a population of 23 candidate coherent structures in two star-forming regions in nearby molecular clouds, L1688 in Ophiuchus and B18 in Taurus, using observations of NH$_3$ emission from the GBT Ammonia Survey \citep{GAS_DR1} and column density maps derived from Herschel observations of dust emission \citep{Andre_2010}.  These cores are identified by a 
 sharp transition from supersonic to subsonic line widths, which determines their boundaries, and a coherent, subsonic non-thermal velocity dispersion in their interiors.  To identify coherent cores, \citet{Chen_2019a} adopt a five-step process, similar to \citet{Pineda_2010}. First, they define the structure boundary as the contour where the thermal and non-thermal components are equal, and each is required to contain a column density peak and local minimum in dust temperature as defined by Herschel. Any region containing multiple NH$_3$ peaks is sub-divided using the emission saddle point. The cores are required to have a signal-to-noise ratio greater than 10 and pixels that produce a large local high-velocity gradient are excluded. 18 of the 23 structures identified by \citet{Chen_2019a} satisfy all five criteria and are considered droplets. The remaining five do not satisfy all the criteria and are therefore considered {\it droplet candidates.}
 The median mass of all 23 cores is $0.2^{+0.3}_{-0.1}$ M$_\odot$, and the median radius is $0.033_{-0.008}^{+0.01}$ pc.  \citet{Chen_2019a} found that the cores have a typical total velocity dispersion, $\sigma_\mathrm{tot} = 0.23_{-0.02}^{+0.01}$ km s$^{-1}$, where
 \begin{equation}
     \sigma_\mathrm{tot} = \sqrt{\sigma_\mathrm{turb}^2 + \sigma_\mathrm{therm}^2}, \label{vdisp}
 \end{equation}
$\sigma_\mathrm{turb}$ is the turbulent velocity dispersion and $\sigma_\mathrm{therm}$ is the thermal velocity dispersion.  These cores have density profiles shallower than a critical Bonnor-Ebert sphere, and they are not bound by self-gravity but are instead bound by pressure provided by the ambient gas motion, i.e., the turbulent pressure.
 
\subsubsection{Pressure-Confined Cores}
\citet{Kirk_2017a} survey dense cores in the Orion A star-forming region. They use gas temperature and velocity dispersion data from GAS \citep{GAS_DR1} and derive core masses and sizes from the James Clerk Maxwell Telescope Gould Belt Survey \citep[JCMT GBS][]{WardThompson_2007}. The JCMT GBS observed 6.2 square degrees around the Orion A molecular cloud at 850 $\mu$m and 450 $\mu$m with SCUBA-2 with resolutions of 14.6\arcsec and 9.8\arcsec.  \citet{Kirk_2017a} adopt the dense core catalogue presented in \citet{Lane_2016}. \citet{Lane_2016} use {\it getsources},  a multi-scale, multi-wavelength source extraction algorithm, to compute the sizes, total fluxes, and peak positions of the cores. {\it Getsources} decomposes the dust emission at each wavelength into a variety of scales and then creates a Gaussian model for the sources, separating them from the surrounding larger-scale emission features \citep{Men_2012}. \citet{Kirk_2017a} approximate the core radii as the geometric mean of the major and minor axis full-width half-max (FWHM) of the {\it getsources} fit and apply a correction for the telescope beam.

The \citet{Kirk_2017a} sample contains 237 cores, of which 26 are cross-matched with {\it Spitzer} sources and classified as protostellar. \citet{Kirk_2017a} find that in fact very few of these cores are sufficiently massive to be bound when considering only the balance between self-gravity and thermal plus internal turbulent motions. This would naively imply that these cores are in the process of dispersing or are non-star-forming. However, the cores are considered bound when the additional pressure imposed by the weight of the ambient molecular cloud is included, suggesting that most of the cores are in fact pressure confined. 

In addition to being a more clustered, higher pressure high-mass star-forming region, gas in Orion is warmer. For the purpose of comparing more directly with our simulated cores, we exclude all observed cores with gas temperatures $\geq$ 15~K, since they have a significantly larger thermal line width then the cores in our simulation. The median mass and radius of the 43 cold dense cores are $0.8_{-0.4}^{+0.3}$ M$_\odot$ and $0.026_{-0.005}^{+0.01}$ pc, respectively.  They have a median total velocity dispersion, $\sigma_\mathrm{tot} =0.32_{-0.04}^{+0.02}$ km s$^{-1}$. 

\subsubsection{Starless Cores in Low-Mass Star-Forming Regions}

\citet{Kerr_2019} present an analysis of starless dense cores identified in three nearby low-mass star-forming regions: Ophiuchus, NGC 1333 in Perseus, and B18 in Taurus. They adopt the same procedure followed by \citet{Kirk_2017a} to identify cores in the JCMT GBS data, combine the footprints with the GAS NH$_3$ data to compute core properties and then estimate the ambient cloud weight from {\it Planck} and {\it Herschel}-based column density maps.

The combined sample totals 132 cores, all starless by construction. Ophiuchus and Perseus also include regions with warmer gas, so as above we exclude all cores in these regions with $T\geq$15~K in the comparison with the simulation data. This leaves a total of 30 cores in Ophiuchus, 33 cores in Perseus and all 8 cores in Taurus. The median mass and radius of the  71 cold dense cores are $0.4_{-0.3}^{+0.4}$\msun and $0.023_{-0.003}^{+0.008}$ pc, respectively.  They have a median total velocity dispersion, $\sigma_\mathrm{tot} = 
$ = $0.37_{-0.05}^{+0.09}$. 

\subsubsection{Virialized Cores in Cepheus}\label{cepheus}

\citet{Keown_2019} analyze the GAS observations of Cepheus-L1251 to identify hierarchical gas structures. To circumvent the complex hyperfine structure of NH$_3$, they construct a simulated Gaussian emission data cube, in which the NH$_3$ structure is represented by Gaussians (the hyperfine structure is effectively removed). They apply {\it astrodendro} to the simulated data to identify 22 high-level structures or leaves, which are equivalent to cores for our purposes. The effective radius of each structure is the geometric mean of the major and minor axes returned by the dendrogram analysis. \citet{Keown_2019} estimate the masses of the ammonia-identified structures using the H$_2$ column density measured by {\it Herschel} dust continuum observations \citep{DiFrancesco_2020}. 

In contrast to the analyses above, \citet{Keown_2019} find that all the cores are roughly virialized, i.e., have comparable kinetic and gravitational energies, without accounting for the contribution of the cloud weight. All of the cores have temperatures below 15\,K, so we include all cores in our simulation comparison. The median mass and radius of the Cepheus-L1251 core sample are $2.5_{-0.8}^{+1.9}$ M$_\odot$ and $0.022_{-0.007}^{+0.005} $ pc, respectively.  They have a median total velocity dispersion, $\sigma_\mathrm{tot} = 
0.23_{-0.01}^{+0.05}$. While the measured sizes and velocity dispersions are similar to those above, the core masses are significantly higher.

\section{Analysis}
\label{sec:analysis}
\begin{figure*}
\includegraphics[width=1.25\columnwidth]{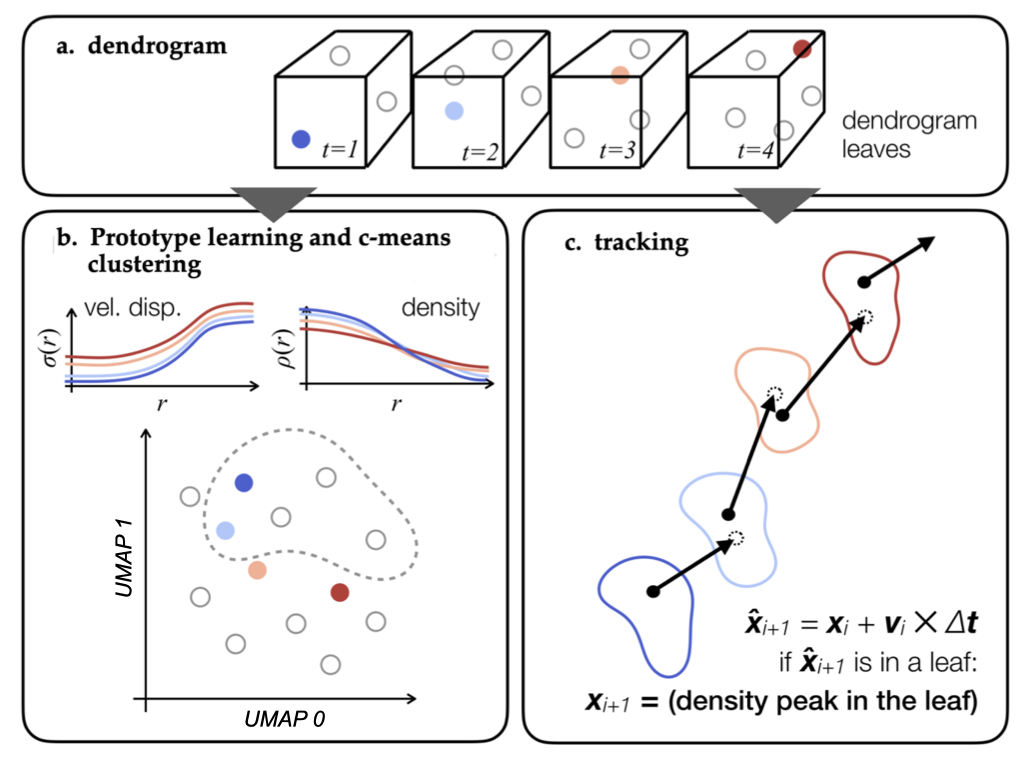}
\caption{  Schematic summary of the analyses carried out in this work.  \textbf{(a)} Density structure identification using dendrograms.  \textbf{(b)} Prototype UMAP analysis and Fuzzy \textit{c}-means clustering analysis on the density profiles, velocity dispersion profiles and core properties.  \textbf{(c)} Tracking each density structure as it moves and evolves across the simulation.  Note that the clustering analysis and tracking are done independently from each other.}\label{fig:summary_analysis} 
\end{figure*}

To carry out a comprehensive analysis of independent density structures in the MHD simulation, we first identify structures using a source extraction algorithm like the one implemented by \citet{Rosolowsky_2008b}, which places structures into a hierarchy as described by a tree-like dendrogram.\footnote{We use \textit{astrodendro}, a \textit{Python} package to extract extended sources in astronomical data (\url{http://dendrograms.org}).} This algorithm is functionally a watershed decomposition algorithm.  We next classify  and visualize the structures using a UMAP and a Fuzzy $c$-means analysis of their properties.
Finally, we track each independent structure in the dendrogram as it evolves and moves across both the simulation and the UMAP space.  Fig.\ \ref{fig:summary_analysis} is a schematic summary of our analysis procedure.

\subsection{Core Identification \& Tracking}
\label{sec:analysis_id}
\begin{figure*}
\includegraphics[width=1.75\columnwidth]{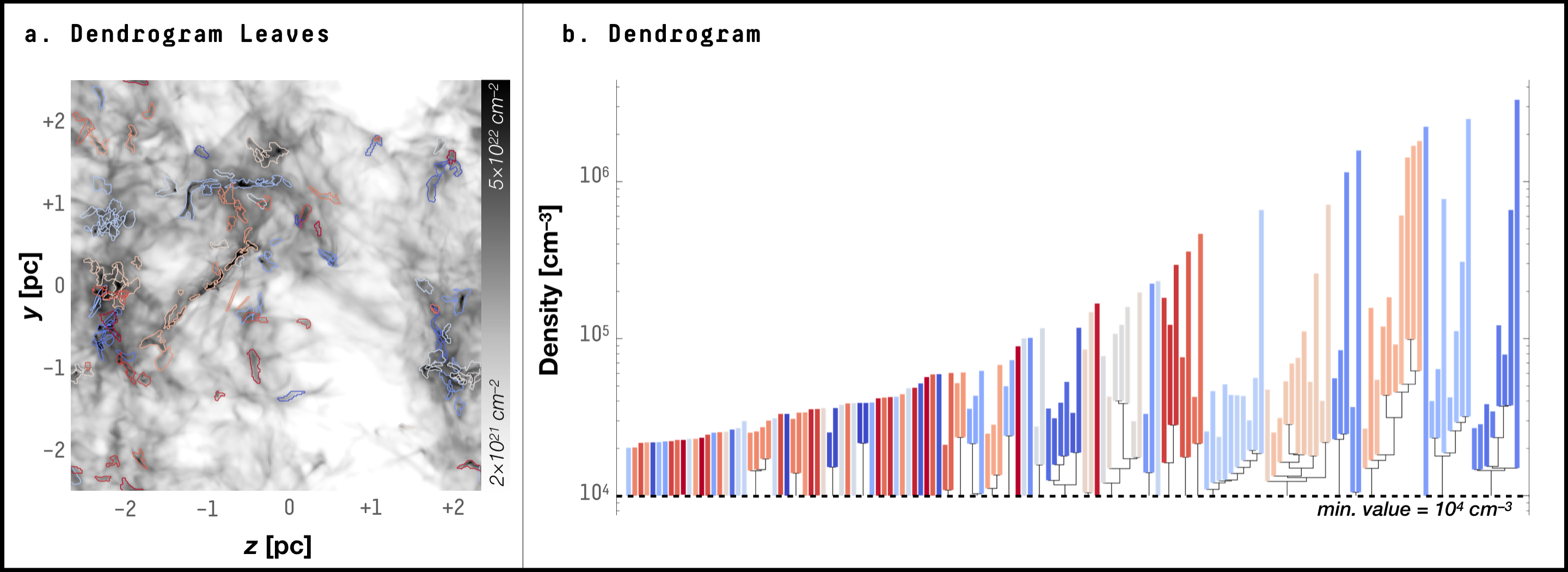}
\caption{Cores identified as dendrogram leaves.  \textbf{(a)} Dendrogram structures plotted on top of the density field integrated over the $x$-axis.  The contours are color coded according to the ID number the \textit{astrodendro} package assigns, and each corresponds to the structure in the dendrogram with the same color.  \textbf{(b)} Dendrogram with the leaves color coded by the ID number the \textit{astrodendro} package assigns.  This snapshot is at $t$ = 4.7$\times$10$^{5}$ yr.  Note that since neighboring structures in the dendrogram are usually assigned consecutive ID numbers, structures that share the same branch may have a difference in color too subtle to be recognized by eye.}\label{fig:leaves}
\end{figure*}

We identify cores in each snapshot of the MHD simulation described in \S\ref{sec:data_simulation} using the 
\textit{dendrogram} algorithm
\citep[hierarchical structure extraction algorithm;
][]{Rosolowsky_2008b, Goodman_2009a}.  Dendrogram-based extraction algorithms (hereafter the dendrogram, for simplicity) efficiently identify density structures in star-forming regions in both simulations \citep[e.g.,][]{Hopkins_2012, Burkhart_2013, Koch_2017} and observations \citep[e.g.,][]{Goodman_2009a, Lee_2014, Seo_2015}.  For each snapshot, we apply the dendrogram on the density distribution in the 3-d space.  We construct the dendrogram to find structures with densities above 10$^4$ cm$^{-3}$, which is characteristic of the densities traced by NH$_3$. To guarantee enough sampling points for the analysis of density and velocity distributions, a structure must have a volume of at least 100 voxels ($\sim$0.02 pc in linear size)  to be included in the dendrogram.  To avoid the inclusion of insignificant local density fluctuations, a structure must also have a difference of 10$^4$ cm$^{-3}$ in density between its peak and the node where it merges onto the tree.\footnote{These setup parameters translate to \textit{min\_value} of 10$^4$ cm$^{-3}$, \textit{min\_delta} of 10$^4$ cm$^{-3}$ and \textit{min\_npix} of 100 in \textit{astrodendro}  A {\it tree} is a full dendrogram representation of hierarchical structures.}.  We identify a total of  3,538 structures over a time span of $6.0\times 10^5$ years, with a nominal time resolution of $\sim 1.5\times 10^4$ years.  Note that we use the dendrogram only to identify independent density structures and locate their peaks.  We do not limit our following analysis of the density distribution to only the density range above 10$^4$ cm$^{-3}$ (see \S\ref{sec:analysis_datadescription} for details), and we only use the dendrogram boundary to avoid confusion with a neighboring core.  See Fig.\ \ref{fig:leaves} for an example of the independent structures identified using the dendrogram algorithm.

To follow the identified cores as they move and evolve in the simulated box, we devise a tracking procedure by first identifying the density peaks within independent structures, \textit{leaves}, in the dendrogram of each snapshot.  The tracking procedure then uses the velocity at the position of the density peak to predict where the density peak is expected to be in the previous and following snapshots.  If the expected position falls within the boundary of a dendrogram leaf, the tracking procedure links the original structure with the leaf in the previous or following snapshot.  This tracking procedure is similar to but less detailed than the one deployed and analyzed by \citet{Smullen_2020}, in which the overlap in various physical quantities and statistical measurements are examined when dendrogram structures in different snapshots are compared.  Our tracking procedure then repeats the process by going through the total of 3,627 independent structures of the dendrograms derived for the snapshots used in this study.  

We find that  3,538 out of 3,627 structures ($\sim$97\%)  are connected to 450 tracks, which link cores identified in two or more snapshots.  As \citet{Smullen_2020} have pointed out, the robustness of the identification using the dendrogram algorithm is subject to uncertainties due to the stochastic fluctuation in the density distribution over time, even when the dendrograms are derived using the same set of input parameters.  We try to avoid the issue of density fluctuations affecting the robustness of dendrogram tracking by excluding structures that are not connected to any of the tracks.  This is equivalent to removing structures that are captured by a dendrogram only in a certain snapshot but not the preceding nor the subsequent ones (separated by $\Delta t\sim$1.5$\times$10$^4$ yr; see above).

Of the 450 tracks, 146 (32\%) end after merging with another track such that they no longer have a unique, distinct peak that can be identified. Since we are particularly interested in the evolution of cores from formation to either star formation or dispersal, we limit our evolutionary study to consider only the 304 main tracks, i.e., we exclude short-lived over-densities that merge with larger ones. We exclude only the minor structure in the merger for the following reasons. If the peak of a structure disappears due to a merger, its track terminates abruptly after a significant jump in the core properties (because the track is matched to a new peak/object). Neglecting these histories allows a cleaner analysis and clearer visualization of evolutionary trends. We, however, include the dominant structure in the analysis since the merger does not abruptly affect the inner profiles near the peak or the bulk properties, which are generally derived from a compact region around the peak.

The average lifetime of the 304 tracks is  $2.15\times 10^5$ years.
 21 tracks span the entire simulation calculation of $\sim 6 \times 10^5$ yr. 
15 out of the remaining 304 tracks ($\sim$5\%) are connected to at least one structure with a sink particle of a mass $\geq$ 0.1 M$_\odot$; several of these are matched to two or three sink particles. 167 of 304 ($\sim$55\% or $\sim 37$\% of 450) cores disperse, i.e., their track ends before forming a sink particle, merging with another track or reaching the last snapshot. Generally, this occurs if the core size or density maximum falls below the dendrogram structure requirement.

\subsection{Constructing Physical Properties of Identified Cores}
\label{sec:analysis_datadescription}

\begin{figure}
\includegraphics[width=0.9\columnwidth]{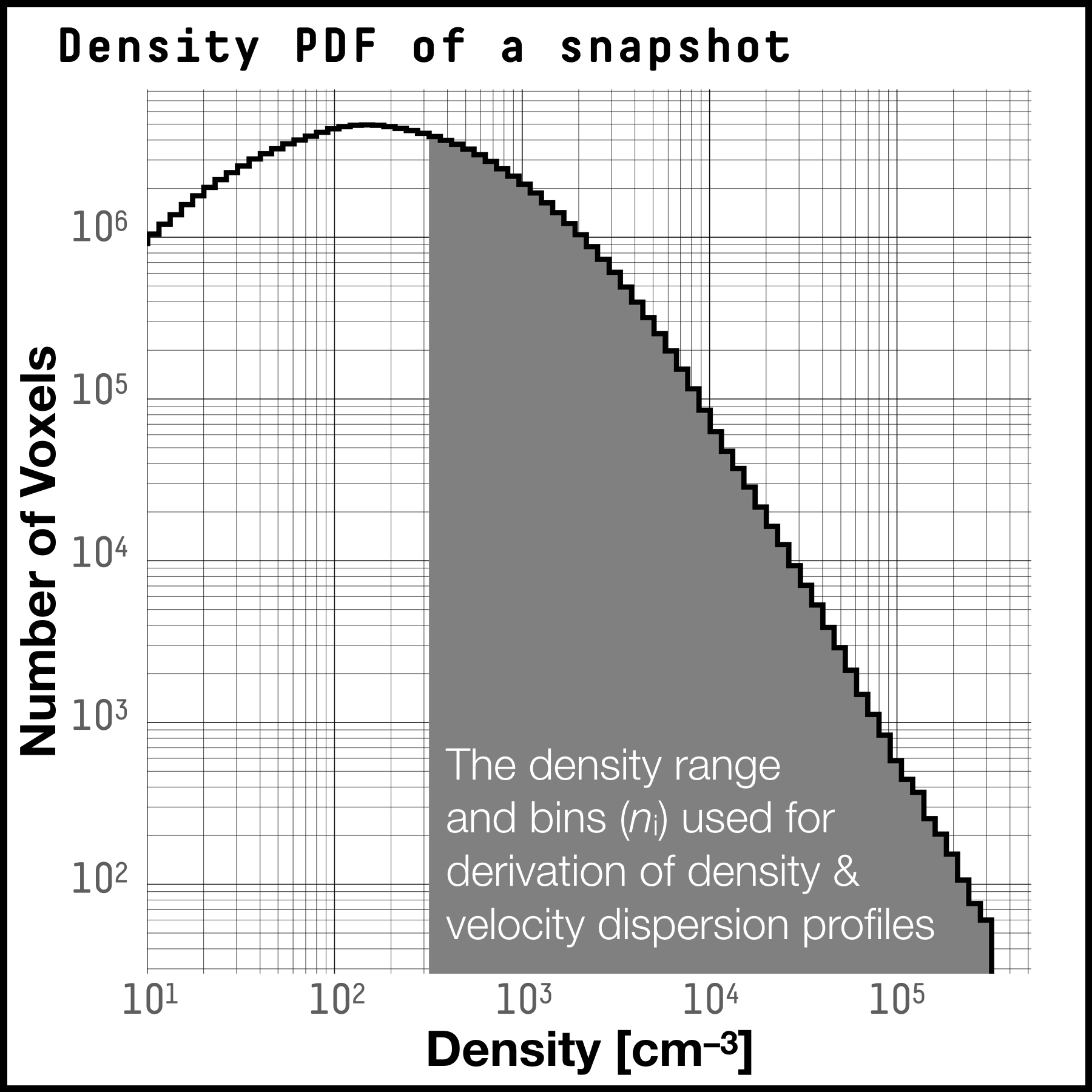}
\caption{ Probability density function (PDF) of density of a snapshot taken at $t$ = 5$\times$10$^5$ yr (solid black line).  The shaded area and bins correspond to the range of density and the series of $n_i$ used for deriving the density and velocity dispersion profiles (see \S\ref{sec:results_phases}).} \label{fig:PDF} 
\end{figure}

In order to analyze the core evolution and compare with observations, we must define a set of fundamental core properties that represent essential characteristics of each core. This step serves as an initial layer of dimensionality reduction, where we reduce the high-dimensional simulation phase space of gas position ($\mathbf{x_i}$), velocity ($\mathbf{v_i}$), and density ($\rho(x_i)$) to a smaller set of parameters that more directly represents each core and can readily be compared with observations.

We first describe each core as a vector of $d=106$ physical properties that contains the radial density and velocity dispersion profiles (50 radial measurements for each), exponent of a power-law fit to the density profile, and bulk core properties, including radius, mass, velocity dispersion, and ratio of kinetic energy to gravitational energy.  We adopt this particular set of bulk properties because they correspond to the set of physical properties previously derived from the observed dense cores in our observational samples (see \S\ref{sec:data_catalogs}). Here, we describe how we derive each of these parameters.

We take the following steps to derive radial profiles.  First, we draw a series of constant density isosurfaces, each at a number density $n_i$.  Since the isosurfaces may take any shape as dictated by the gas distribution, we make no assumption about the geometry of the cores.  We use 51 density values uniformly spaced on a logarithmic scale from $n$ = 10$^{2.5}$ cm$^{-3}$ to 10$^{5.5}$ cm$^{-3}$. As Fig.\ \ref{fig:PDF} shows, these densities sample the underlying probability density function (PDF) of gas density well.  Each isosurface is then converted to an equivalent radius by finding the radius that would construct a sphere that has the same volume as the volume enclosed by the isosurface, i.e., $V_\mathrm{iso} = 4 \pi R_\mathrm{eq}^3/3$.\footnote{ We note this definition is the 3-d equivalent of the effective radius that is often derived in observations of clouds and cores \citep{Rosolowsky_2006}.} The radial density profile, $n(r)$, is then constructed from the series of densities, $n_i$, that define the isosurfaces and the corresponding equivalent radii, $R_\mathrm{eq, i}$.  For the velocity dispersion profile, we calculate the velocity dispersion of material enclosed within each isosurface, $\sigma_i$, and similarly construct the profile of velocity dispersion, $\sigma(r)$, from $\sigma_i$ and $R_\mathrm{eq, i}$.  Note that the profile represents the 3-d turbulent velocity dispersion and does not include the thermal sound speed. The structure boundaries defined by the dendrogram are only used to avoid confusion with another core. We stop the construction of profiles when the volume enclosed by the isosurface overlaps with the dendrogram boundary of another core.  This occurs mostly when the core has a \textit{sibling}, i.e., a nearby leaf that has the same density minimum and shares the same \textit{parent} branch in the dendrogram.  For a core that does not have a sibling \citep[the trunk-leaves---independent structures at the bottom level;][]{Rosolowsky_2008b}, the  extent of the radial profile is not limited by the dendrogram structure boundary (see \S\ref{sec:analysis_id}).
This method does not involve spherical averaging and can produce radial profiles for structures with different shapes in a reliable and consistent way.

We use the 1-d profiles to derive the rest of the core properties. In order to better compare with the observations described in \S\ref{sec:data_catalogs}, we define the boundary such that the core radius, $R_c$, is the FWHM of the density profile. This definition is similar to that adopted by the {\it getsources} algorithm, which is commonly used to define observed structures. While this does not allow a true apples-to-apples comparison, using the FWHM as the core boundary produces simulated core with masses, sizes and velocity dispersions comparable to the those of observed cores (see \S\ref{sec:obs_comparison}). We derive the core mass, $M_c$, by integrating the density profile to obtain the mass enclosed by $R_c$. Since observations do not include protostellar information in core estimates, we exclude the sink mass in the calculation of $M_c$ and all the other core properties.
For the total velocity dispersion of the core, we adopt the observational definition in Equation \ref{vdisp}. Here, $\sigma_{\rm turb} = \sigma(R_c)/\sqrt{3}$ and $c_s$ is the sound speed for a 10~K molecular gas. We define the radius of coherence, $R_{\rm coh}$, as the radius where the velocity dispersion falls below the sound speed: $\sigma(r)/\sqrt{3} < c_{\rm s}$. We obtain the density power-law index by performing a least squares fit on the density profile for $r<0.1$~pc.

Using the mass, the size and the velocity dispersion, we derive the kinetic energy and the gravitational potential energy. For the purpose of later observational comparison (see \S\ref{sec:obs_comparison}), we adopt the expressions from \citet{Chen_2019a}, where the kinetic energy is
\begin{equation}
\Omega_{\rm K} = \frac{3}{2}M_c\sigma_{\rm tot}^2
\end{equation}
and the gravitational energy is
\begin{equation}
\Omega_{\rm G} = -\frac{3}{5}\frac{GM_c}{R_c}. \label{eq:grav}
\end{equation}
The latter expression assumes the cores have a uniform density distribution. 
Cores with a density profile $\rho \propto r^{-2}$ will have an actual gravitational energy a factor of $\sim1.7$ times larger than that expressed in Equation \ref{eq:grav} \citep{Pattle_2015}.  

To evaluate the impact of the choice of core definition on our analysis, we also adopt a fixed density contour to define core boundaries. We present this analysis in Appendix \ref{sec:appendix_coredef}. There we demonstrate that while the quantitative distribution of core properties depends on core definition,  the qualitative determination of phases and our conclusions are reasonably robust to the core definition.

After deriving the properties for each core, we remove duplicate information by reducing the number of profile data points that contribute to the final data vector. We describe our procedure in Appendix \ref{sec:appendix_lasso}. This effectively reduces the weight of the profiles in the later analysis, so that the bulk and profile information is considered more equally. This process reduces the 100 profile values to 22.

Finally, we assemble a data matrix composed of  $d=28$ physical property measurements for each of the $N = 3,538$ structures identified by the method of \S \ref{sec:analysis_id}.

\subsection{Core Clustering Methodologies}
\label{sec:analysis_methodoverview}

\begin{figure} \centering
\includegraphics[width=0.97\columnwidth]{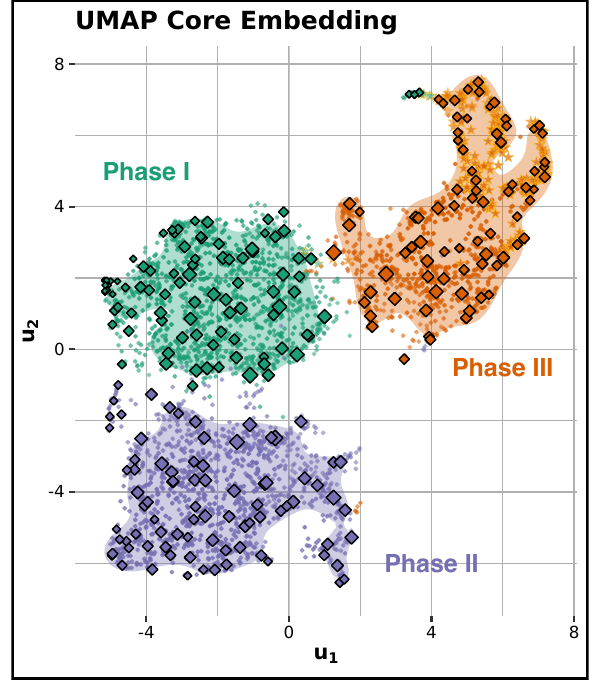}
\caption{  A two-dimensional UMAP Embedding (using 36 neighbors, see Appendix \ref{sec:appendix_umap}) of the 3,538 cores identified from the simulation (points) and 249 neural gas prototypes learned from them (diamonds). Colors indicate cluster (phase) membership, while their transparency represents their cluster membership strength $U$ (fainter points belong less confidently to their reported cluster); both are determined by the FCM algorithm applied to the high-dimensional core profiles. Prototype sizes are mapped to the number of cores each represents, which is determined during a recall of the entire training dataset through the neural gas network. Shading indicates a 75\% highest density region of a phase-conditional kernel density estimate fit to the embedded points, which is shown to facilitate cluster boundaries in UMAP space. Stars indicate sink particles identified from simulation.} 
\label{fig:tsne_embedding} 
\end{figure}

 Our goal is to identify groupings of the 3,538 cores in order to differentiate
evolutionary behavior based on physical properties. Because our data arise from discrete snapshots of the continuous process of an MHD simulation (\S \ref{sec:data_simulation}) we have reason to suspect the boundaries separating (defining) each phase are less crisp than those arising from a truly discrete process. This complicates the clustering task, whose goal is delineation of such boundaries. To aid cluster saliency while still acknowledging the fuzziness of our data groupings we employ two approaches from unsupervised machine learning: (1) we learn prototype representations of our data and then (2) create a soft partitioning of these prototypes based on the Fuzzy $c$-means algorithm. The benefits of this two-pronged approach are discussed in the next two sections.

\subsubsection{Learning Prototypes of Core Properties}
\label{sec:neural_gas}

 Prototype-based methods in machine learning \citep{biehl2016prototype} apply common machine learning tasks (e.g., clustering or classification) to intelligently formed representations of the data called \textit{prototypes} (instead of the data themselves).  That is, from $N$ data observations $X = \{x_i \in \mathbb{R}^d \}_{i=1}^N$ we learn $M$ prototypes $W = \{ w_j \in \mathbb{R}^d \}_{j=1}^M$. The prototypes arise from the codebook of a vector quantizer \citep{Gray1984} trained on $X$ and benefit the learning task by simultaneously reducing sample size (typically $M << N$) and decreasing noise (the process of quantizing an $x_i$ by its best representative $w_j$ separates the signal and noise components of $x_i$). While classical $k$-means \citep{macqueen1967} with a large number of centroids is a common method for obtaining prototypes, in this work we obtain $M=249$ 
 prototypes of our $N=3,538$ cores from the Batch Neural Gas algorithm (\citealt{cottrell2006batch}, extended from \citealt{MartinetzSchulten1991}) trained on the core properties.  Neural vector quantizers (Neural Gas, as well as the Self-Organizing Map, see \citealt{kohonen2001}) benefit from a cooperative element during their training process, rendering them less sensitive to the initialization issues common for $k$-means \citep{cottrell2006batch}. No theory currently exists for selecting an optimal number of prototypes: there should be enough to fully capture intricacies of the data distribution, but not so many that the vector quantizer approaches an identity mapping. Often, analyses adopt empirical rules of thumb from related areas such as kernel density estimation that suggest $M  = \mathcal{O}(\sqrt{N})$.  Here we select the optimal number via an iterative process: we start with M=100 and learn. If all prototypes are utilized, i.e., there are no dead prototypes with empty receptive fields, we increase $M$ by 50 and repeat, stopping once the set contains at least one dead prototype. This process yielded $M$ = 249 prototypes (1 dead prototype was removed after the last iteration) for the 3,538 cores; because we find a similar number of clusters and set of cluster properties for $M=150$, we conclude that our analysis does not strongly depend on the number of prototypes within a factor of 2.

Beyond sample size and noise reduction, vector quantization provides a unique prototype similarity measure which we consult for intelligent parameterization of part of our analysis (see Appendix \ref{sec:appendix_umap} for details).  For completeness we also compared our results with those derived from a more basic principle component analysis (PCA) and from a self-organizing map (SOM) of the core data. We find that both these approaches return qualitatively similar cluster organization and cluster assignments. We present the Neural Gas prototype analysis here, since it provides the best combination of group separation and simplicity.

\subsubsection{Fuzzy c-means Clustering}
\label{sec:cmeans}

 Once learned, the core prototypes are clustered by a user-selected method and the cores themselves inherit the cluster label of their best representative. The continuous nature of our data (\S \ref{sec:analysis_methodoverview}) suggests we should expect some cluster overlap; to account for this, we choose a soft partitioning of the core prototypes by the Fuzzy $c$-means algorithm (or FCM, \citealt{bezdek1984fcm}). Typical hard partitioning schemes assume well separated data clusters and, consequently, assign data to a single cluster. Soft partitionings instead report a membership strength $U_{ik}$ representing the degree to which datum $x_i$ belongs to cluster $k$. By convention, $0 \leq U_{ik} \leq 1$, $\sum_k U_{ik} = 1$, where $U_{ik} > 0.5$ denotes a datum's strong membership in cluster $k$. Importantly, the graded information contained in $U$ influences the formation of cluster centers in soft partitioning algorithms. For completeness, we note that hard partitionings are a special case of soft partitionings where the $U_{ik}$ are constrained to the set $\{0,1\}$. From the analysis of Appendix \ref{sec:appendix_cluster}, FCM applied to our core prototypes suggests $c=3$ clusters (evolutionary phases) exist in the simulated core sample. To mitigate initialization issues, the clusterings reported in this work are optimal, i.e., have lowest within-group error over 1,000 different randomly initialized runs of FCM.

\subsection{Visualization with UMAP}
\label{sec:umap}

 Note that the evolutionary tracks described in \S \ref{sec:analysis_id} were {\it not} used by FCM during the clustering procedure; therefore, the resulting partitioning produces clusters of cores with similar physical properties. Our goal is to uncover a relationship between these groupings and a core's evolution. To this end we employ a 2-d visualization of core prototypes via the UMAP algorithm \citep{mcinnes2018}, which serves two purposes: 1) it allows inspection of the integrity of the three FCM-identified clusters and 2) provides an organized space upon which to view the core tracks.  Figure \ref{fig:tsne_embedding} shows the  UMAP visualization of the prototype data and the resulting three clusters identified as described in \S\ref{sec:cmeans}.  UMAP has gained popularity relative to other common approaches for dimensionality reduction, such as t-SNE, due to its visualization quality, ability to retain high-d structure in the lower-d projection and calculation speed. The data visualizations (e.g., Figures \ref{fig:tsne_embedding} and \ref{fig:tSNE_var_projections}), along with associated group-wise statistics of Figure \ref{fig:clusterprops} and Table \ref{table:properties} underpin the evolutionary interpretation of our clustering, as discussed in \S \ref{sec:results_evolution}.  An overview of  UMAP and an explanation of the parameters used in this work can be found in Appendix \ref{sec:appendix_umap}.

\begin{figure}
\begin{center}
\includegraphics[width=0.95\columnwidth]{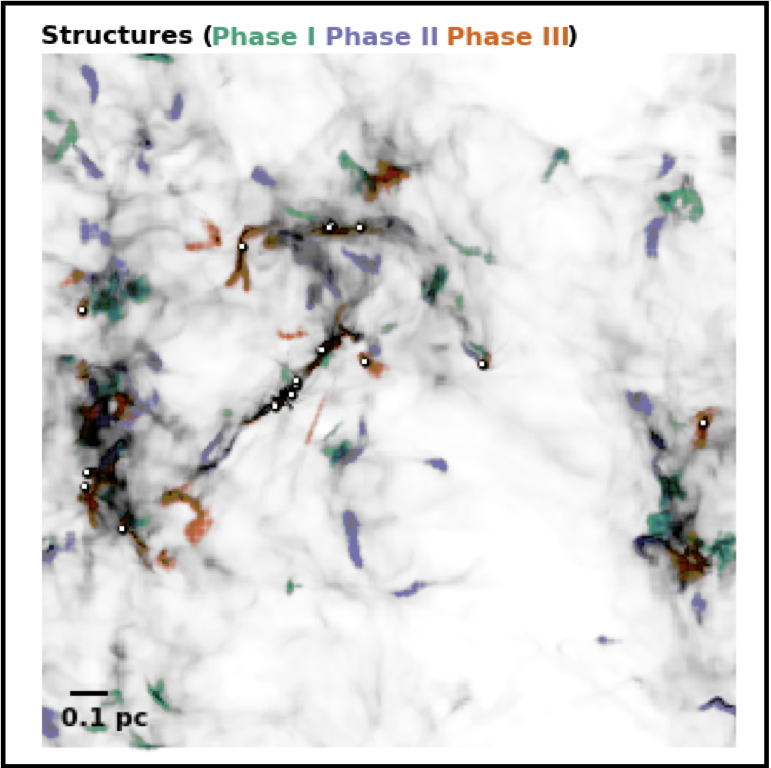} 
\caption{ Structures at $4.7\times 10^5$ yr overlaid on the gas column density and colored by their assigned phase. White dots indicate the location of sink particles. The time and view are the same as in Figure \ref{fig:leaves}.}\label{fig:dendro_phase}
\end{center}
\end{figure}

\section{Results}
\label{sec:results}

\begin{table*}
    \setlength\tabcolsep{2.0pt} 
	\centering
	\begin{tabular}{ || l | c | c | c | c | c  | c | c |c | c | c | }
	\hline
    Core Classification & $N$ &  $M_c$~(\msun) & $R_c$~(pc) & $R_{\rm coh}$~(pc) & p & $\sigma_\mathrm{tot}$~(km s$^{-1}$) & $V_{\rm bulk, 1d}$~(km s$^{-1}$) &     $\Omega_\mathrm{K}$/$\left|\Omega_\mathrm{G}\right|$  & $f_{\rm *}$\,(\%) & $\bar d$\,(pc) \\ 
	\hline
	Phase I (Turbulent) &  1221 & 0.3$_{-0.1}^{+0.2}$ & 0.034$_{-0.008}^{+0.008}$ & 0.012$_{-0.004}^{+0.004}$ & -0.9$_{-0.2}^{+0.2}$ &  0.27$_{-0.02}^{+0.03}$ & 0.6$_{-0.2}^{+0.2}$  &  6.3$_{-1.6}^{+3.1}$ &  0.82 & 0.17$_{-0.07}^{+0.11}$\\ 
	Phase II (Coherent) &  1317 & 0.4$_{-0.1}^{+0.2}$ &  0.040$_{-0.008}^{+0.007}$ & 0.029$_{-0.006}^{+0.009}$ & -0.9$_{-0.2}^{+0.1}$ &  0.23$_{-0.01}^{+0.02}$ &0.4$_{-0.2}^{+0.3}$  & 3.2$_{-0.6}^{+0.9}$ & 0.0 &  0.18$_{-0.07}^{+0.13}$  \\
	Phase III (Protostellar) &  1000 & 0.3$_{-0.2}^{+0.2}$ & 0.023$_{-0.004}^{+0.006}$ &  0.008$_{-0.008}^{+0.007}$ &  -1.2$_{-0.3}^{+0.2}$ &  0.26$_{-0.02}^{+0.04}$ & 0.6$_{-0.2}^{+0.2}$ & 2.9$_{-1.0}^{+1.6}$ & 22.9 & 0.13$_{-0.05}^{+0.06}$ \\ 
	\hline
	All & 3538 & 0.3$_{-0.1}^{+0.2}$ & 0.032$_{-0.008}^{+0.01}$ & 0.016$_{-0.007}^{+0.01}$ & -0.9$_{-0.3}^{+0.2}$ &  0.25$_{-0.02}^{+0.03}$ & 0.5$_{-0.2}^{+0.3}$ & 3.9$_{-1.1}^{+2.1}$ & 6.8 & 0.16$_{-0.06}^{+0.10}$  \\ 
	\hline
    \end{tabular}
        \vspace{-0.1cm}
 \caption{  Physical properties of cores in each phase. We assign those that have partial membership in two different clusters to the one with the highest membership. The physical properties are measured using the density and velocity profiles derived from the dendrogram structure. The columns are number of cores and median core mass, radius, size of the coherent region, density index, total velocity dispersion, bulk velocity, ratio between the kinetic energy and the absolute value of the gravitational potential energy, fraction of members containing protostars and nearest neighbor separation. The density index is the power-law index of the function, $n = n_0 (r/r_0)^p$, fitted to the density profile of each core. The spreads are calculated using the 0.25 and 0.75 quantiles of the distribution.
 }
 \label{table:properties}\vspace{-0.5cm}
\end{table*}

\subsection{Properties of Core Phases}
\label{sec:results_phases}

 Table \ref{table:properties} summarizes the simulation core properties for all 3,538 cores and for cores classified in each of the phases. While the core masses are similar across all phases, clear differences appear in the other median properties. Phase I and Phase II cores have similar masses, sizes and density indices, however Phase II cores contain a large subregion with a subsonic non-thermal velocity dispersion, i.e., a region of coherence \citep{Pineda_2015,Chen_2019a}. Consequently, we term Phase II the {\it coherent} phase.  Phase II cores also have a slightly lower overall non-thermal dispersion and a lower bulk velocity. Phase III cores have the steepest density index ($p = -1.2_{-0.3}^{0.2}$) and the lowest ratio of kinetic to gravitational energy ($\Omega_{\rm K}/|\Omega_{\rm G}|=2.9_{-1.0}^{+1.6}$). Since our calculation for the gravitational potential assumes a uniform potential these virial parameters are likely over-estimated by a factor of 1.7, which means that most of the Phase III cores are gravitationally bound. We also find  $\sim 23$\% of these contain sink particles (compared to $0.8$\% and 0\% of Phase I and II cores, respectively). Therefore, we term Phase III the {\it prestellar/protostellar} phase.   Of the three phases, Phase I has the highest ratio of kinetic to gravitational energy.  Consequently, we refer to Phase I as the {\it turbulent} phase.  In order for cores in this phase to form stars they must either gain significant mass or reduce their gas velocity dispersion  (possibly  by passing through Phase II). 

Cores almost always belong to Phase III after forming protostars (see Figure \ref{fig:tsne_embedding}), so it can be loosely considered the last phase. However, there is no one  evolutionary order between I, II and III and not all cores that belong to Phase III at a given time go on to form protostars (see \S\ref{sec:results_evolution}  for more discussion). Cores may form in any phase and take a variety of different routes to evolve through the parameter space until they become protostellar or disperse as we discuss in detail in \S\ref{sec:results_evolution}. 

Figure \ref{fig:dendro_phase} shows a column density map with the identified structures colored by their phase. Most of the Phase III cores are located within large filaments, which is also where most of the protostars reside. Many of the Phase I and II structures are associated with shocks and/or more isolated filamentary features. They also tend to be larger and have lower column densities, which is consistent with being gravitationally unbound.

\begin{figure*} \centering 
\includegraphics[width=1.99\columnwidth]{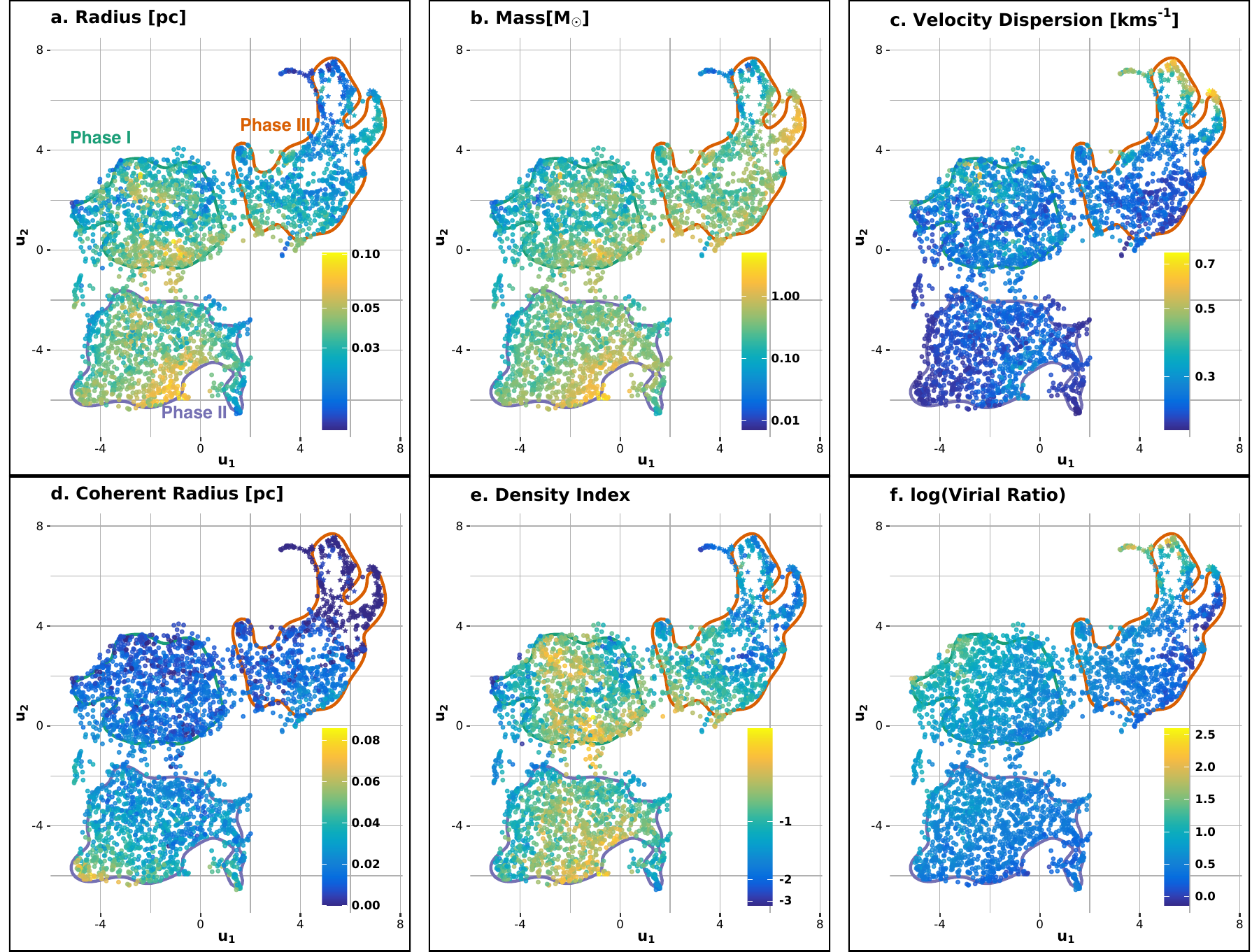}
\caption{  
 Projection of  six different core properties (\textbf{(a)} radius, \textbf{(b)} mass, \textbf{(c)} velocity dispersion, \textbf{(d)} radius of coherence, \textbf{(e)} density index, \textbf{(e)} virial ratio) to the embedded core locations in UMAP space.  
}
\label{fig:tSNE_var_projections}
\end{figure*}

Figure \ref{fig:tSNE_var_projections} shows the distributions of core radii, masses, velocity dispersion, virial ratio (ratio of kinetic to gravitational energy),  density index and size of the coherent region. The clusters do not divide cleanly across any of these properties, but there is evidence of property gradients. For example, Figure \ref{fig:tSNE_var_projections}a shows core sizes transition from large to small from bottom to top. The core mass distribution exhibits similar structure as shown in Figure \ref{fig:tSNE_var_projections}b, with the lowest mass cores appearing at the top of Phase I and Phase III.
 Similarly, Figure \ref{fig:tSNE_var_projections}c shows a strong vertical gradient in velocity dispersion, which is echoed in the distribution of virial ratios shown in Figure \ref{fig:tSNE_var_projections}f.
There are two distinct regions of high virial ratio: one appears in Phase I, where cores seem to be genuinely unbound due to high levels of turbulence, and the other occurs in the topmost corner of Phase III, where the high dispersion is produced by infall. The prototypes within the  lower region of Phase III have the lowest virial ratios, suggesting that cores are becoming bound as they approach the stage of gravitational collapse.    Unlike the others the density index exhibits stronger horizontal gradients, with steeper profiles on the very right and left, while flatter profiles appear in the center. Figure \ref{fig:tSNE_var_projections}d shows the cleanest and most monotonic trend across phases of all six properties: there is a strong vertical gradient in the size of the coherent region, with the most coherent cores located at the bottom left of the UMAP (Phase II) and cores with no coherent region at the top right (Phase III).

\begin{figure*}
\centering 
\includegraphics[width = 1.95\columnwidth]{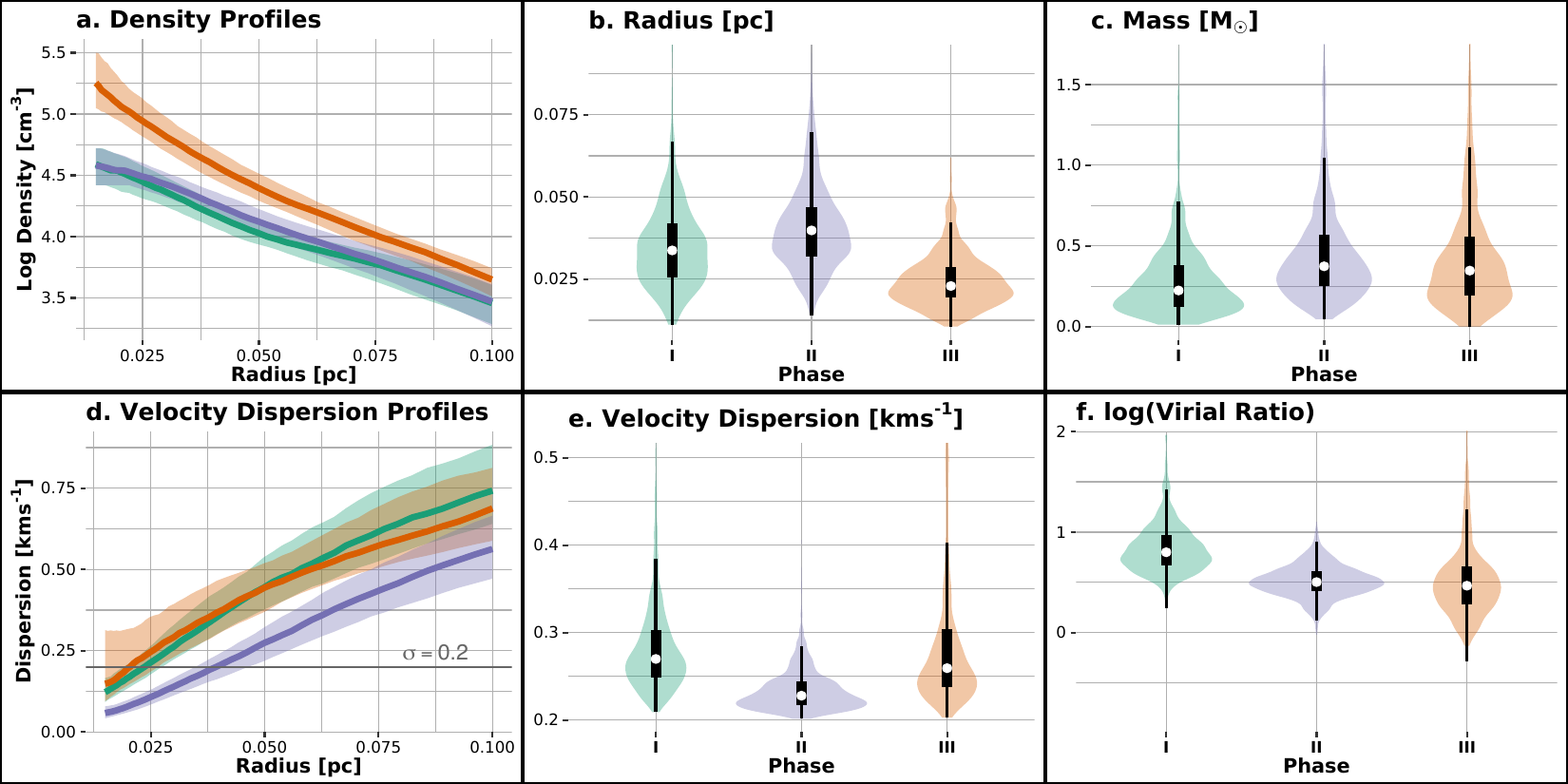}
\caption{  Summary of cluster statistics. Radial profiles of density \textbf{(a)} and 3-d velocity dispersion \textbf{(d)} for each of the three clusters, where thick lines represent the median profile and the spread is the interquartile range. The horizontal grey line in \textbf{(d)} denotes the value at which the turbulent velocity dispersion equals the sonic speed at 10~K. The violin plots show the distributions of intra-cluster \textbf{(b)} radius, \textbf{(c)} mass, \textbf{(e)} velocity dispersion and \textbf{(f)} virial ratio. The interquartile range (thick black lines), median (white point) and Tukey's fences (thin black lines) have been added to the violin plots to aid cluster comparison.  
}
\label{fig:clusterprops}
\end{figure*}

 Fig.\ \ref{fig:clusterprops} displays the density and non-thermal velocity dispersion profiles for each of the clusters (left panels) and the distributions for radius, total velocity dispersion, mass and virial ratio (center and right panels).  With the exception of mass,  the profiles and properties exhibit distinct differences for the three phases. Phase I and II have significant overlap in several of the properties but are distinguished by the velocity dispersion: Phase I cores are more turbulent  at all radii, while Phase II cores have velocity dispersion profiles that dip to sub-sonic values near the core center, i.e., they have an internal coherent region. This difference in velocity dispersion is also reflected by the virial ratio, which tends to be higher for Phase I cores.
Phase III cores exhibit noticeably steeper density profiles with a higher central density.  Meanwhile, the velocity dispersion of Phase III cores is typically supersonic  for all radii with velocity dispersion  flattening or increasing near the center.  This feature, together with the steeper density profile, is consistent with gravitational infall dominating the internal kinematics of the core and the incipient formation of protostars. For this reason, Phase III cores are also more compact on average because the FWHM corresponds to a smaller region (see Appendix \ref{sec:appendix_coredef}).

\subsection{Core Evolution}
\label{sec:results_evolution}

In this section we use the core histories and cluster assignments to explore how cores evolve through the cluster phase space. 

We first calculate how long cores typically spend in each of the three phases. By averaging over the time cores spend visiting each phase, we derive an effective phase lifetime; cores that never visit a phase are not included in its time average.  We estimate typical lifetimes of 
 1.0$\pm$0.1$\times$10$^5$ yr, 
1.3$\pm$0.2$\times$10$^5$ yr, and 
1.8$\pm$0.3$\times$10$^5$ yr
for Phase I, II and III, respectively. We find that a core evolving into Phase III spends significantly longer there. For example, cores that eventually form protostars spend 
0.6$\pm$0.3$\times$10$^5$ yr visiting Phase I and/or II and 
5.0$\pm$0.4$\times$10$^5$ yr in Phase III.
This is because star-forming cores remain in Phase III after becoming protostellar and also because the lifetimes of cores that visit Phase III tend to be systematically longer. The lifetime of Phase I is the shortest, which is consistent with most of the cores being unbound.

\begin{figure*}
    \centering
    \includegraphics[width=1.99\columnwidth]{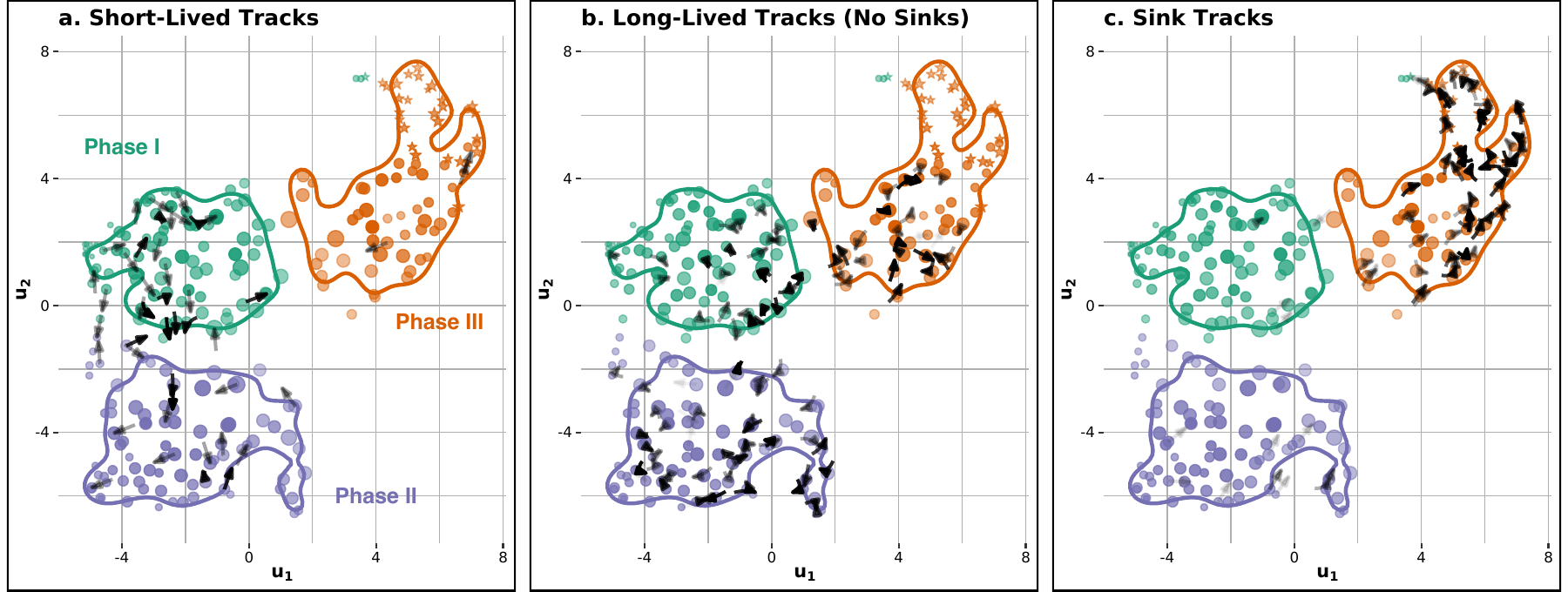}
    \caption{ Directional evolution of cores following short-lived \textbf{(a)}, long-lived \textbf{(b)}, and sink tracks \textbf{(c)}. Short-lived tracks exist in only 2 of the 26 time snapshots of the MHD simulation, long-lived tracks persist throughout, and sink tracks contain cores that form protostars at some point during their duration. Arrows were constructed by a B\'ezier fit using the following control points in UMAP space: median direction \textit{from} which cores transition to each prototype (arrow tail), the prototype itself (middle), and the median direction \textit{to} which cores transit after visiting each prototype (arrow head). (Shorter) arrow length indicates  (mis-)alignment of the incoming / outgoing directions. Short-lived cores are predominantly mapped to Phase I and II and star-forming cores migrate into Phase III, while long-lived tracks inhabit the middle of the diagram and cross through all three phases. 75\% Highest Density Regions of the clusters are outlined by color, and arrow transparency represents the number of tracks forming their direction.}
    \label{fig:evolution_clusters}
\end{figure*}

Next we investigate the trajectories of cores through the phase space. Figure \ref{fig:evolution_clusters} shows tracks for three different sets of core histories: {\it short-lived tracks}, which connect cores that appear only in two snapshots, {\it long-lived tracks}, in which the cores persist for all simulation snapshots but do not form stars, and {\it sink tracks}, which represent the evolution of cores that eventually become protostellar.
Arrows represent the aggregate direction of movement for all cores passing through the associated prototype, constructed as a quadratic B\'ezier curve with control points set by the median incoming direction (arrow tail), the prototype itself, and the median outgoing direction (arrow head). 
The unit vectors describing the incoming/outgoing control points are further scaled by the proportion of  incoming/outgoing tracks transiting through each prototype. Thus, higher arrow curvature indicates more misalignment between the median incoming and outgoing track directions, and an asymmetry in arrow length (relative to the arrow's middle elbow) indicates areas of core birth (longer outgoing head) or dissipation (longer incoming tail). 

As UMAP is a highly non-linear manifold projection, some of the strong curvature observed in Figure \ref{fig:evolution_clusters} is to be expected. For example, prototypes representing sink particles appear in a circular region in the top right as shown by Figure \ref{fig:evolution_clusters}c, and the arrows connecting neighboring prototypes naturally possess curvature to follow the circular structure in an organized manner.  However, in more linear regions of the embedding, curvature indicates track reversal of the incoming / outgoing movement of a prototype's typical core. The strongest examples of such core meandering occur in the long lived tracks of Fig. \ref{fig:evolution_clusters}b, indicating that these tracks bounce from one prototype to another (i.e., they migrate between different set of physical characteristics) continuously due to small changes in their properties.  One fundamental implication of this figure is that there is no one evolutionary path for cores.

The short-lived tracks represent relatively transient cores that quickly disperse. These tracks inhabit the top left part of the phase space, lying almost entirely within Phase I and II. Many of the arrows point 
 along the edge or outwards as if they are exiting the  UMAP boundaries. These cores disappear because their densities and/or sizes fall below the threshold of detection by our dendrogram algorithm, which is consistent with the small masses and sizes of cores in this region of the parameter space (e.g., compare Fig.~\ref{fig:evolution_clusters}a and Fig.~ \ref{fig:tSNE_var_projections}ab). 

The long-lived tracks inhabit the middle of the UMAP, spanning parts of Phase I, II and III. They appear to complement the short-lived tracks, since their motion is concentrated in the right half of Phase I and the bottom of Phase II. Their longevity suggests that they have achieved some degree of equilibrium, and inspection of many of these cores indicates that they become coherent, moving into Phase II, and remain there for much of their lifetime. This is illustrated by the shortness of the arrows, which indicate that many cores mapped to prototypes in the middle of Phase I and II do not undergo rapid or significant changes in their properties between snapshots. The general impression is that this subset of cores evolve more gradually between phases. 
Since there is no preferred phase where cores start, the initial position is not predictive of the longevity or the direction of evolution.

The behavior of the cores following sink tracks is potentially the most interesting, since these cores are the subset that eventually form stars. The arrow directions generally point towards the upper right, suggesting that these cores move rightwards and upwards in the parameter space as they evolve. Cores with sink particles lie almost exclusively in the top right corner of Phase III, which is consistent with the apparent trajectory of these cores. Prestellar cores, i.e., those that later go on to form stars, mostly ( 9 of 15) start in Phase II. These cores become protostellar while in Phase III, in a region of the parameter space in which the virial ratio is small, and remain in Phase III for the remainder of their evolution. Despite spending most of their evolution in Phase III, 
73\%
of cores that eventually become protostellar spend time in another Phase: on average 
0.6$\pm$0.3$\times$10$^5$ yr visiting Phase I and/or II and 
5.0$\pm$0.4$\times$10$^5$ yr in Phase III.
Note that  prototype locations in Phase III can also host some short and long-lived cores, and thus the initial core properties and phase space location are not entirely predictive of the eventual evolution. 

\begin{figure*}
    \centering
   \includegraphics[width=1.99\columnwidth]{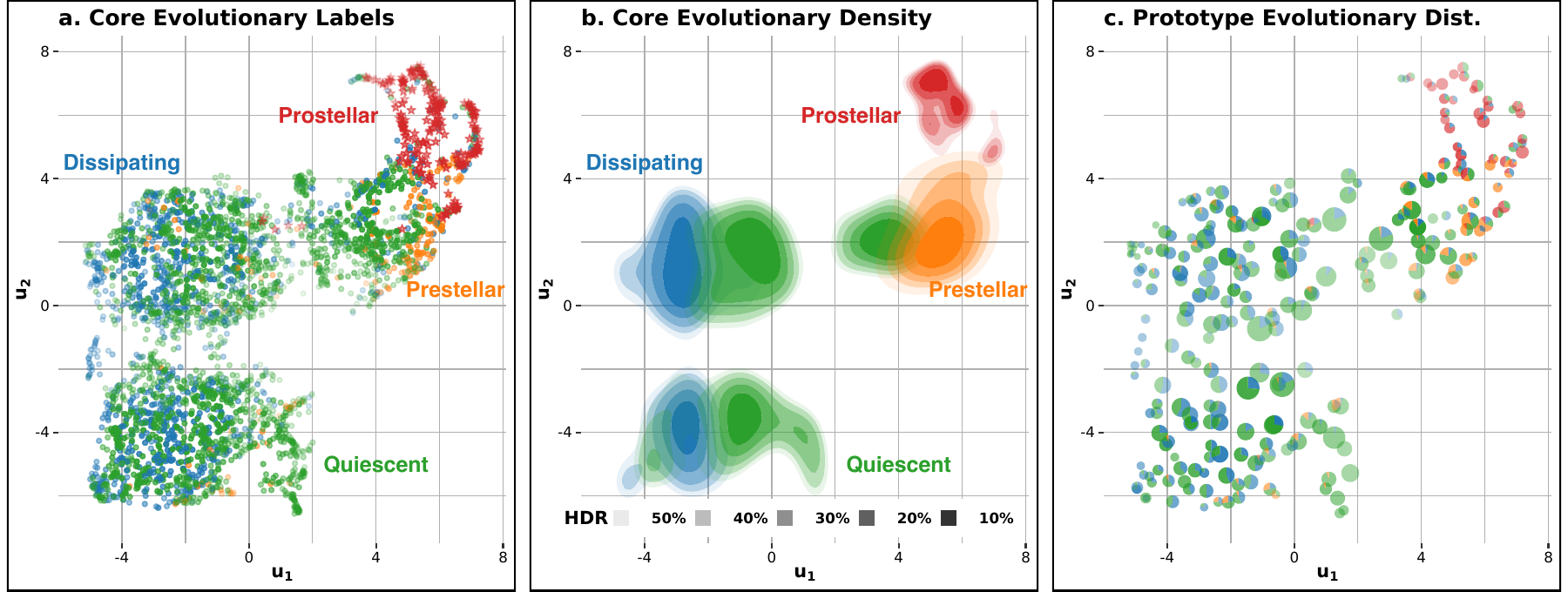}
    
 \caption{  Core evolutionary labels projected to UMAP space. 
 \textbf{(a)} Cores are colored by their evolutionary status, as observed at the end of simulation. From left to right, UMAP  organizes evolution into dissipating, quiescent, and pre/protostellar regions. Note that the evolutionary organization bleeds across the cluster footprints identified by FCM in Fig. \ref{fig:tsne_embedding}, indicating that no single association exists between properties and evolutionary status. For comparison, point transparency represents the strength $U$ of corresponding cluster membership from \S  \ref{sec:cmeans}. 
 \textbf{(b)} 10-50\% \textbf{H}ighest \textbf{D}ensity \textbf{R}egions of a 2-d Kernel Density Estimate of cores in UMAP space, conditioned on evolutionary status. A visualization such as this could provide probabilistic prediction of the evolutionary fate of any observed cores (e.g., those presented in Figure \ref{fig:observed_cores}), although we have yet to formally classify such. 
 \textbf{(c)} Pie charts placed at the neural gas prototype locations in UMAP space display the distribution of evolutionary labels in each prototype's receptive field (or RF, which is the set of points mapped to them). Size corresponds to the cardinality of each prototype's RF, while transparency indicates the prototype's cluster membership strength $U$.}
\label{fig:evolution_projection}
\end{figure*}

Finally, in Figure \ref{fig:evolution_projection} we synthesize the evolutionary information by coloring the UMAP not by cluster membership but by the outcome of the evolution of the cores passing through each prototype. Here we denote four states: cores that are protostellar (red), cores that are prestellar and will eventually become protostellar (orange), cores that disperse (blue) and cores that neither disperse nor form protostars by the end of the simulation (green). Many cores comprising the last class have reached a quasi-equilibrium state due to magnetic, turbulent and thermal pressure support, and they are represented by the long-lived tracks. Figure \ref{fig:evolution_projection}c shows that prototypes on the left represent predominantly dispersing cores, while pre/protostellar cores are almost exclusively mapped to prototypes on the right.

Note that the core histories are not included in the information used to perform the clustering, and thus represent an independent view of how the clusters relate to one another. In many cases, the clustering appears to intuit some of the evolutionary movement, since related prototypes, e.g., those representing star-forming cores, are confined to specific regions of the visualization.  However, Figure \ref{fig:evolution_projection} shows the evolution is not cleanly represented by particular properties, which show a vertical separation. While cores assigned to Phase I and Phase II have distinct properties, these properties only partially predict whether the cores will disperse, persist or form stars  (see \S\ref{sec:results_timescales} for further discussion of evolutionary rates). Disorder in the UMAP is produced by the stochastic nature of star formation: core properties vary as a function of the environment, formation  and accretion history. 

\begin{figure*}\centering
\includegraphics[width=0.45\linewidth]{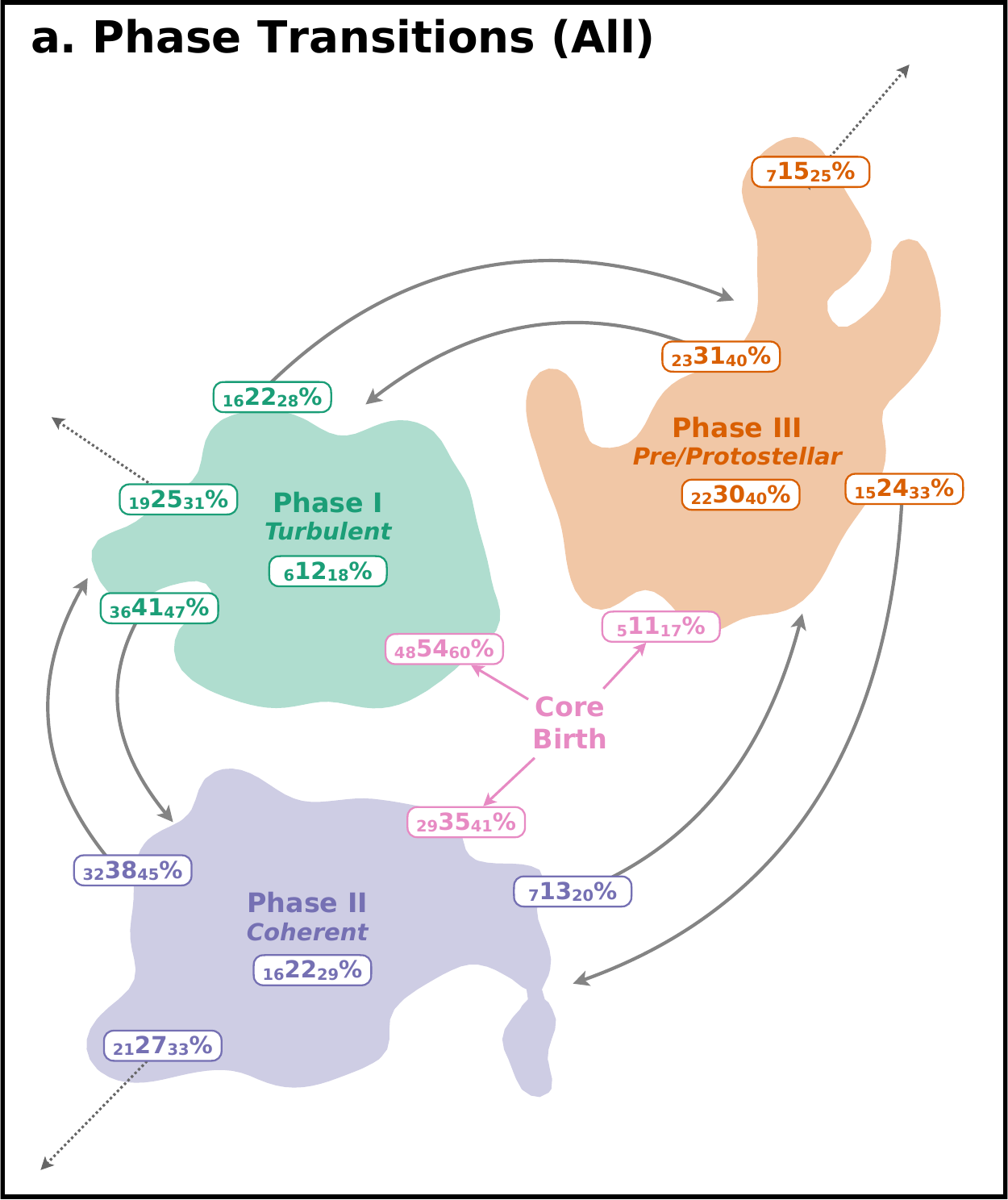} \hspace{0.03\linewidth}%
\includegraphics[width=0.45\linewidth]{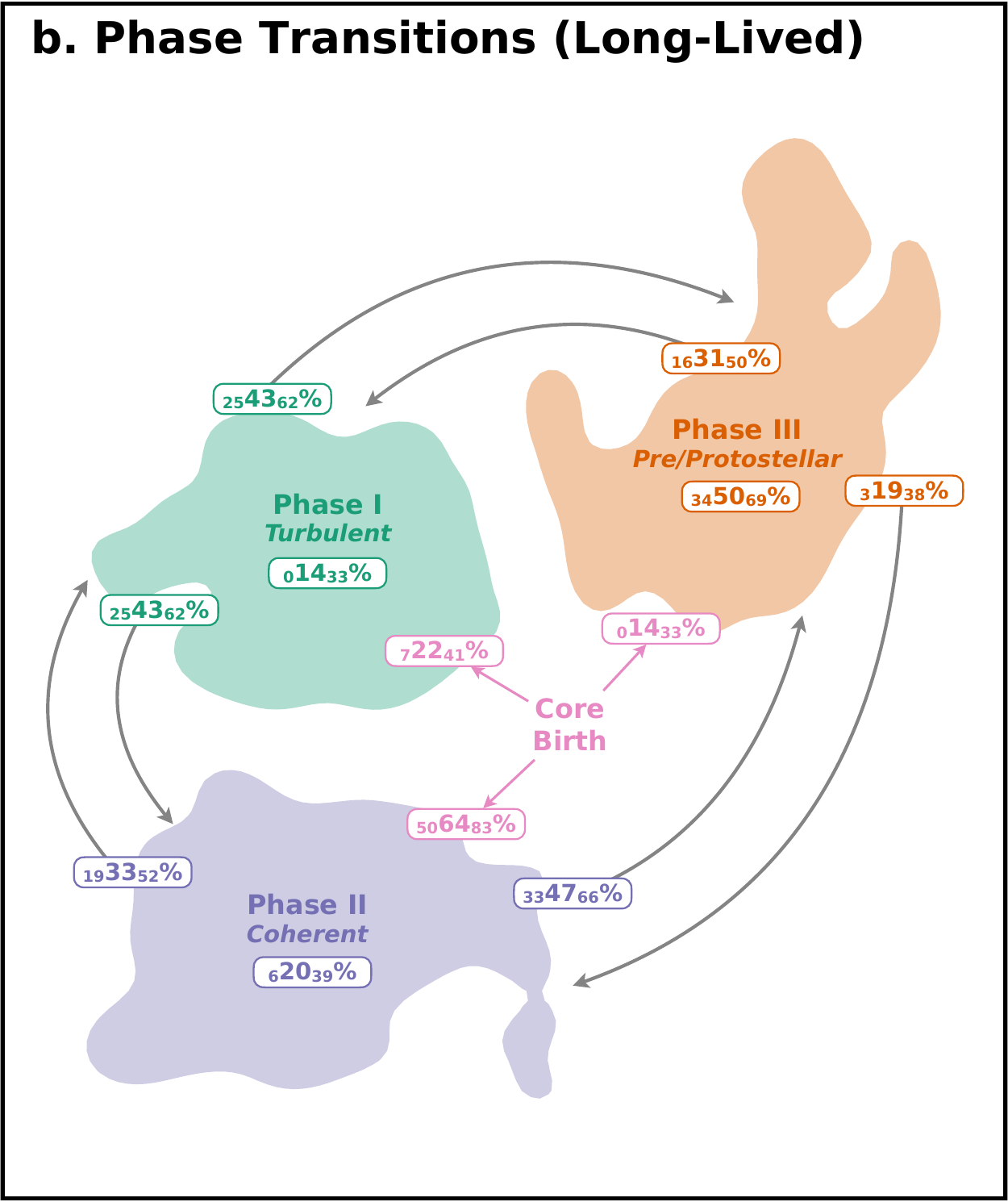}%

\caption{ 
 A summary of the transition probabilities among the three phases, as estimated empirically from the track histories, visualized in the organized UMAP space (with axes removed for readability). The percentage ($\pm$ 95\% multinomial confidence intervals, \citealt{glaz1999simultaneous}) of core transitions from one cluster to another is shown at the start of directed paths connecting each cluster (larger boxed text), while the percentage of stationary cores is displayed immediately below the cluster name.  The dashed paths leaving each cluster represent core dispersal, which we consider to be another state space for transition.   Pink text indicates the percentage of cores which first appear ({\it are born}) in each cluster. All point estimates of the same color add to 100\%. Panel \textbf{(a)} computes these percentages relative to all tracks,  while panel \textbf{(b)} considers only long-lived  (appearing in every snapshot of our simulation) and  protostellar tracks. 
} 
\label{fig:phase_evolution}
\end{figure*}

\subsection{Survival Rates \& Lifetimes}
\label{sec:results_timescales}

In \S\ref{sec:results_evolution}, we show that evolutionary tracks exist that connect three populations of cores with different physical properties.   A closer examination of the survival rates, defined as the fraction of cores remaining in a given phase, reveals that cores classified in the same phase can follow distinctly different evolutionary paths. Figure \ref{fig:phase_evolution} shows the percentages of cores in a given phase that stay in that phase, eventually move to another phase and/or disperse. For example, if a core starts in Phase I, moves into Phase II, and then moves to Phase III before finally dispersing, it will be counted in the statistics of cores that  are born in Phase I (54\%), move from I to II (41\%), move from II to III (13\%) and then disperse from III (15\%).  If that core belongs to the subset of cores that persist or eventually form stars Figure \ref{fig:phase_evolution}b shows that for this path only 22\% are born in Phase I,  43\% move from I to II and  47\%  move from II to III.  Stated another way, this figure shows the transition probabilities for a core observed in a given phase. For example, if a core is currently observed in Phase III, the probabilities of either transitioning next to I or II or to dispersing from Phase III are shown in the figure. We include 95\% confidence intervals to give a sense of the uncertainties based on the core statistics.

We find that all cores have a relatively high probability of phase transition: 
85$\pm$4\%
either move to another phase, disperse, or both, during the simulation, while 
 55$\pm$6\%
of cores belong to two or more phases during their evolution.
 Phase I cores are most transient with only   $12_{6}^{18}$\% chance that a core in that phase remains there for 
 the remainder of its life. Approximately a quarter of the cores disperse from each phase, with cores in Phase II having the lowest survival rate and Phase III cores having the highest (only $15_{7}^{25}$\% cores disperse from this phase).

Figure \ref{fig:phase_evolution}a shows there is a lot of movement between Phase I and II. While it is most likely that a Phase I core transitions into Phase II  $41_{36}^{47}$\%, there is a nearly equal probability, $38_{32}^{45}$\%, of a Phase II core transitioning to Phase I ( see also Fig.~\ref{fig:evolution_clusters}).
Phase III cores are most likely to remain in their current phase, in part because 23\% of Phase III cores are protostellar. Phase III cores that do leave are more likely to move into Phase I ($31_{23}^{40}$\%) than into Phase II ($24_{15}^{33}$\%). This core subset has a significant amount of initial turbulence: they can't immediately collapse because they are not bound by gravity.
Figure \ref{fig:phase_evolution} shows that while most cores are born into Phase I (54\%), the majority of cores that persist or form stars, i.e., the ones that don't disperse, begin in Phase II (64\%). In either case very few cores start in Phase III. 

\begin{figure}
\includegraphics[width=0.95\columnwidth]{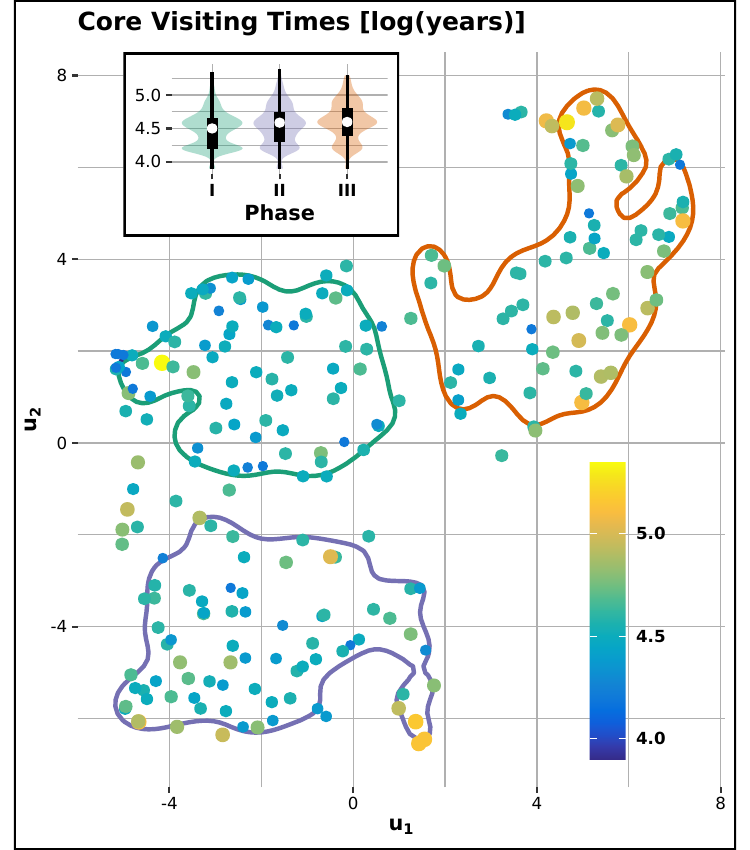}
\caption{  Median time (log(years)) that cores spend visiting (being represented by) each prototype along their evolutionary track, represented in UMAP space. The marker size also corresponds to time. \textbf{Inset:} Distribution of visiting times by evolutionary phase, which can also be considered the prototype "lifetime." Prototypes with longer visiting times, such as those in Phase III, indicate that the core properties are stable and change relatively slowly.
}\label{fig:tsne_timescale}
\end{figure}

Note that while the phases can be described by average properties, there is a range of properties within each phase. This is also illustrated by Figure \ref{fig:tsne_timescale}, which shows the distribution of prototype {\it visiting times,} i.e., how long a typical core is matched to a given prototype. For example, cores in the upper right of Phase III are not likely to change phase or disperse because most already host stars. This is also reflected in the longer time periods a core matches a given prototype in this region. Interestingly, Figure \ref{fig:tsne_timescale} shows there is another grouping of long-lived prototypes towards the bottom of Phase II. Inspection of Figure \ref{fig:clusterprops} indicates that these are moderately-sized cores that are marginally bound and quiescent, i.e., these are coherent cores that have reached a quasi-equilibrium state. In contrast, the prototypes in Phase I tend to have the shortest lifetimes 
(3.8$\pm$0.5$\times$10$^4$ yr vs. 
5.2$\pm$0.6$\times$10$^4$ yr in Phase II and 
5.9$\pm$0.7$\times$10$^4$ yr in Phase III
),
indicating that the properties of Phase I cores change relatively quickly. 

\subsection{Core Properties}
\label{sec:results_properties}
 
 In this section we present an analysis of the physical properties derived using the core profiles constructed from the dendrogram-identified hierarchy.

\begin{figure*}
\includegraphics[width=1.5\columnwidth]{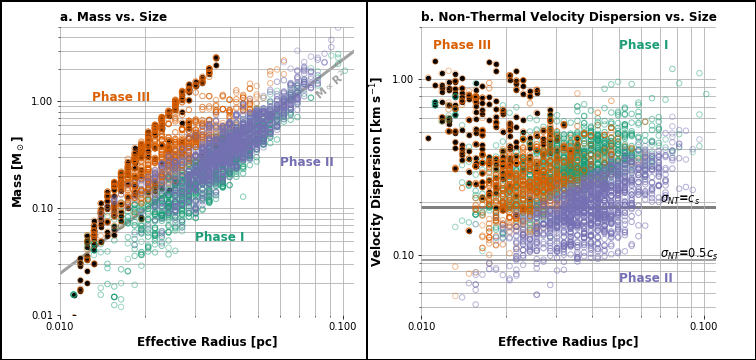}
\caption{\label{fig:properties_groups}  \textbf{(a)}  Mass-size distribution of all 3,538 independent structures.  The green, purple and orange circles correspond to structures in Phase I, II and III, respectively. The symbol transparency is set by the weight of the core cluster assignment. Black filled circles indicate cores with sink particles. The grey line shows a fit to the Phase I and Phase II core populations. \textbf{(b)} Non-thermal velocity dispersion-size distribution of all 3,538 independent structures, with a color coding scheme the same as (a).  The horizontal black lines denotes the velocity dispersion values when the non-thermal velocity dispersion is equal to the sonic speed (thicker line) and half the sonic speed (thinner line) at 10~K. Nearly all protostellar cores are members of Phase III. They tend to be more compact and have higher velocity dispersions compared to other cores.}
\end{figure*}

 Fig.\ \ref{fig:properties_groups}a shows mass as a function of size for cores in each of the three phases.  The phases generally fall along a power-law relation where the Phase III cores, which are often protostellar, are offset to a higher mass at a given radius. The protostellar cores are more centrally peaked such that the FWHM core definition returns more compact structures. 
A power-law fit to the mass-size distribution of cores belonging in all three phases gives a power-law index of $\sim$1.5.  A fit to only the Phase I and Phase II cores returns a power-law index of $\sim2.0$, as expected from Larson's relations \citep{Larson_1981}. Appendix \ref{sec:appendix_coredef} shows that the power-law index is sensitive to the core definition, however.

Fig.\ \ref{fig:properties_groups}b shows non-thermal velocity dispersion, $\sigma_{\rm turb}$, as a function of size for structures in each of the three phases.  As expected from the velocity dispersion profiles examined in \S\ref{sec:results_evolution}, Phase I and Phase III cores generally have larger velocity dispersions than Phase II structures,  which generally have subsonic dispersions. Protostellar cores have the largest velocity dispersions due to gravitational infall. Since the simulations neglect mass-loss due to protostellar outflows, the sink particles are over-massive \citep{Smullen_2020} and the degree of infall, and hence the non-thermal component, is likely over-estimated. 

\begin{figure}
\includegraphics[width=\columnwidth]{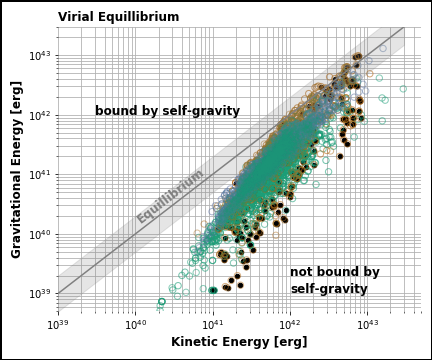}
\caption{  Gravitational potential energy, $|\Omega_{\rm G}|$, versus kinetic energy, $\Omega_{\rm K}$, for all 3,538 structures.  The green, purple and orange circles correspond to structures in Phase I, II and III, respectively.  The red band from the lower left to the top right marks equilibrium between the gravitational potential energy and the internal kinetic energy (grey line) within a factor of two (grey shaded region).}\label{fig:virial_groups}
\end{figure}

Fig.\ \ref{fig:virial_groups} shows gravitational energy versus kinetic energy for cores in the three phases.  Such a comparison, conventionally known as a \textit{virial analysis}, provides a first-order estimate of the gravitational boundedness of a structure.  A virial analysis may sometimes include other terms such as the magnetic energy and the surface pressure term \citep[see][]{WardThompson_2006, Pattle_2015, Chen_2019a}.  Since the core mass does not include the sink mass, we note the gravitational binding energy of the protostellar core is underestimated. 
 We find that there is no clear separation in the distribution of kinetic and gravitational energies between Phases. In contrast, see the analysis in Appendix \ref{sec:appendix_coredef}, which also shows that these properties are sensitive to the core definition. However, there appear to be more Phase III cores with high gravitational and kinetic energy that are more gravitationally bound, consistent with the star-forming activities found within many of them.  Phase I and II cores are almost all below the equilibrium line and are unbound when considering only thermal, gravitational and kinetic energy. Recall that our definition for the gravitational energy in Equation~\ref{eq:grav} assumed a uniform density; we see here this description is more accurate for Phase I and Phase II cores, which have a relatively flat density profile.

\begin{table*}
    \setlength\tabcolsep{2.0pt} 
	\centering
    \begin{tabular}{ || l | c | c | r | c | r | c | r | }
	\hline
	& & \multicolumn{2}{c}{\textbf{Disperse}} & \multicolumn{2}{c}{\textbf{Persist}} & \multicolumn{2}{c}{\textbf{Pre/Protostellar}} \\
    \textbf{Region} & $N$(total) &  $N$ & \% & $N$ & \% & $N$ & \% \\ 
	\hline
 	Ophiuchus & 30 & 6 & 21$\pm$9\% & 4 & 21$\pm$9\% & 20 & 58$\pm$14\%  \\
 	Orion & 43 & 3 & 19$\pm$5\% & 14 & 26$\pm$8\% & 26  & 55$\pm$12\% \\
 	Cepheus & 22 & 0 & 7$\pm$2\% & 2 & 13$\pm$10\% & 20 &  80$\pm$10\% \\
 	Perseus & 33 & 2 & 17$\pm$6\% & 12 & 31$\pm$11\% & 19 & 52$\pm$13\%  \\
     Droplets & 23 & 18 & 54$\pm$12\% & 4 & 39$\pm$11\% & 1 & 7$\pm$7 \\
     Taurus & 8 & 2 & 21$\pm$27\% & 6 & 65$\pm$24\% & 0 & 14$\pm$8\%  \\
 	\hline
	All &  159 & 31 & 22$\pm$4\% & 42 & 28$\pm$4\% & 86 & 50$\pm$6\%   \\
 	\hline
    \end{tabular}
        \vspace{-0.1cm}
 \caption{Predicted future evolution for cores observed in each star-forming region. 
 The table reports the number of observations from each region predicted to be in each evolutionary state along with the mean$\pm$95\% confidence interval of the class-wise predictive probabilities. 
 }
 \label{table:classification_cores}
\end{table*}

\subsection{Classification of Observations}
\label{sec:obs_comparison}

\begin{figure}
\includegraphics[width=0.95\columnwidth]{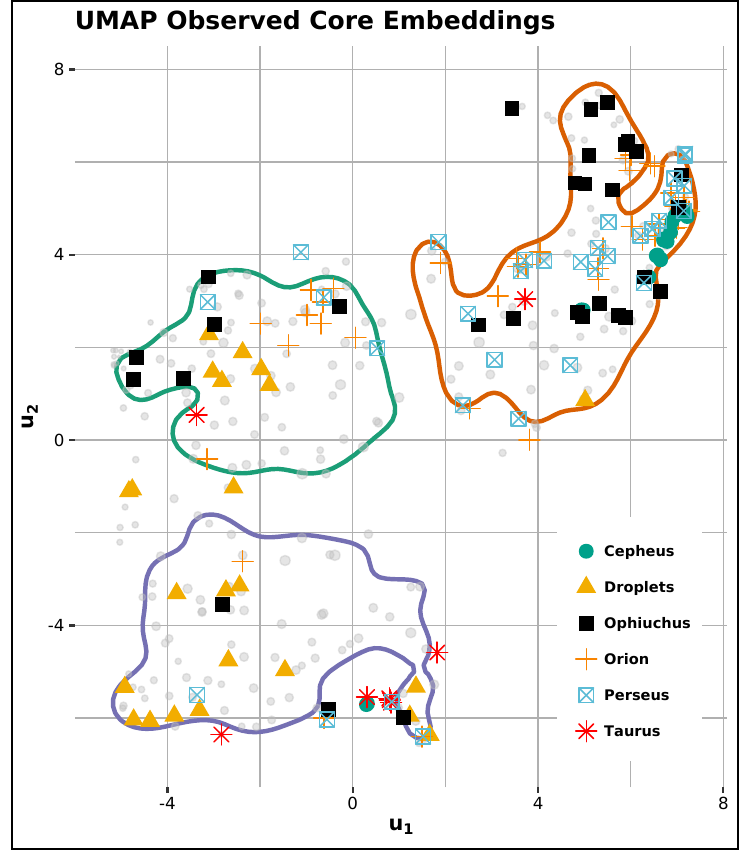}
\caption{ Observed cores (\S \ref{sec:data_catalogs}) embedded in UMAP space according to the procedure outlined in \S \ref{sec:obs_comparison}.  Some prototypes represent multiple observations from our data catalog, but many have no observational match. Conclusions from the analysis relating  cluster-wise physical properties to evolutionary phase (\S \ref{sec:results_phases} ) apply most confidently to observations located inside the outlined 75\% cluster highest density regions. 
}\label{fig:observed_cores}
\end{figure}

 In this section we compare the observed cores with the simulated cores by using their properties to match them to prototypes and project them into the UMAP parameter space. Each observed core inherits coordinates in the UMAP plane from their most representative prototype among those trained on our simulated cores according to \S \ref{sec:neural_gas}. Recall (\S \ref{sec:analysis_datadescription}) that each prototype represents 28 different physical core properties, with the radial density and velocity dispersion profiles comprising 22 of the 28. As this information is missing from the observed cores, we have mapped observations to prototypes based solely on their radius, mass, velocity dispersion, and virial ratio 
by excluding the radial profiles learned by the neural gas prototypes during quantization. The size of the cores identified by \citet{Chen_2019a} is defined by construction to be the size of the coherent region, so we use the coherent radius instead of the radius to project this sample into the UMAP.  We acknowledge that the neural gas algorithm may well have learned to represent this reduced four-dimensional space differently (i.e., produced a different set of prototypes), but any re-training would necessitate a separate clustering (\S\ref{sec:cmeans}) and produce a different UMAP embedding (\S\ref{sec:umap}).

We note that 33 (of 159) observed cores have a property that falls slightly outside the range of the properties of the simulated cores. The Cepheus cores, which adopt a different core definition and appear the most bound of all the core catalogs, have the most discrepancy. However, since these differences are within the observational uncertainties, we do not exclude them from our comparison. Inspection of their phases and location in UMAP space indicates that their classification is still consistent with the expectation given their general properties.

 Figure \ref{fig:observed_cores} shows the observed cores are mapped to locations across the UMAP space. In some cases, multiple cores in different regions are mapped to the same prototype, as in the top right, while other prototypes have no observational match. The droplets identified by \citet{Chen_2019a} are mostly mapped to prototypes in Phase II. This is consistent with droplets being quiescent, coherent structures by definition.  
 The cores observed in Taurus \citep{Kerr_2019} are likewise mostly mapped to prototypes that are classified as Phase II. 

In contrast, few cores in Perseus, Ophiuchus and Orion \citep{Kirk_2017a,Kerr_2019} match prototypes in Phase II. These cores predominantly belong to Phase I or III, and they are instead located in regions of the parameter space characterized by high velocity dispersions and high virial ratios (top of the UMAP) as shown in Figure \ref{fig:tSNE_var_projections}. The Perseus and Ophiuchus cores were selected to be starless by construction, and their correspondence with prototypes in the top right -- where the simulated protostellar cores lie -- may either mean they are prestellar and close to forming stars or that their properties are similar because they belong to more clustered environments, which is also true of the simulated protostellar cores (see Table \ref{table:properties}). The Cepheus cores from \citet{Keown_2019} are all mapped to a few prototypes in the top right of Phase III, a region of the parameter space containing mostly prestellar, bound simulated cores (see Fig.~\ref{fig:tSNE_var_projections}).

To be more quantitative, we employ the simulated core distribution to predict the evolution of the observed cores. Table \ref{table:classification_cores} lists the predicted probabilities of future evolution for cores observed in each star-forming region. 
 To obtain these predictions, we mapped the observations  to their most similar simulated core using the available subset of bulk properties (radius, velocity dispersion, mass, and virial ratio). UMAP coordinates for each observation were inherited from its nearest simulated neighbor, and a Bayesian classification probability was obtained using the kernel density estimates of each evolutionary stage visible in Figure \ref{fig:evolution_projection}b. In most regions about $\sim$20\% of cores are expected to disperse while $\gtrsim 50$\% are expected to eventually form stars. The droplets have the highest rate of expected dispersal ($54$\%), while the Cepheus cores have the lowest ($7$\%). $65$\% of the cores in Taurus are expected to `persist', i.e., they are likely long-lived quasi-equilibrium structures. This is consistent with Taurus being a quiescent region where the   star-formation is most distributed. Overall, at least $\sim20$\% of the observed cores are expected to disperse, while at least half are likely to form stars in the near term.

 Figures \ref{fig:compdata} and \ref{fig:coherent} compare the properties of the individual observed cores to the simulated cores.  As shown by the prototype comparison in Figure \ref{fig:observed_cores}, there is good agreement between properties of observed and simulated cores. In the 2-d parameter spaces of physical properties there is significant overlap between the phases, so it is not always clear which phase an observed core belongs to, for example, on the basis of velocity dispersion and radius, alone. However, we can still infer some general trends by inspecting the distribution of observed core properties.

Figure \ref{fig:compdata}a displays total velocity dispersion versus effective radius for the three phases and the observed cores.  Most of the droplets lie in the Phase II region, which has a lower total velocity dispersion and where the total is dominated by the thermal component. The cores in the warmer and more clustered regions -- Orion, Perseus and Ophiuchus -- lie predominantly in the Phase I and Phase III regions, where the velocity dispersions are higher. By construction most of these cores are starless and relatively few fall into the high-dispersion, compact size region (upper-right Phase III quadrant) where the simulated protostellar cores lie. The Taurus and Cepheus cores generally fall within the Phase I and II regions. As we discussed in \S\ref{sec:results_timescales} the simulations predict a high level of core dispersal, and the location of the observed starless cores in phase space is not predictive of whether a core will {\it definitively} go on to form stars (although cores found in the middle part of Phase III are more likely to be or become star-forming).  

 Figure \ref{fig:compdata}b shows gravitational energy versus kinetic energy for the three phases and the observed cores. There is likewise a high degree of overlap between the phases, which suggests that the virial ratio cannot uniquely determine the core phase.  In this space, there is also good agreement between the simulated and observed cores with most of both appearing to be unbound. However, a subset of the observed cores have high gravitational energies and these extend outside the simulation parameter space. Nearly all of these are cores in Cepheus, which were defined using the {\it dendrogram} leaf boundary and thus are systematically larger than cores in the other clouds.  Our analysis in Appendix \ref{sec:appendix_coredef} suggests that in fact the low virial ratios may be partially due to the core definition.

 Figure \ref{fig:coherent} shows core mass versus coherent region size for the three phases and cores from \citet{Chen_2019a}. This data is only available for the droplet population, which are explicitly identified and defined by the extent of the coherent region.  The droplets fall almost entirely within the simulated Phase II region; two have significantly higher masses and sizes. While there is some overlap between the three phases, the resolution of the observations appears to limit the minimum detected size of the internal coherent subregion, such that any detected sizable coherent region uniquely identifies cores as belonging to Phase II. The simulation phase distributions suggest that other observed cores likely contain coherent regions with sizes below the observational resolution ($\sim 0.02-0.05$ pc).

\section{Discussion}
\label{sec:discussion}

\subsection{Predicting Core Evolution}
\label{sec:discussion_evolution}
 Based on the results presented in \S\ref{sec:results}, we propose an evolutionary scenario where cores inhabit three distinct phases.  Cores in these three phases bear characteristically different physical properties.  
 In summary, cores are born as turbulent density structures that, depending on their initial size and virial ratio, may belong to any of the three phases. A subset of the smallest and most unbound cores quickly disperse (e.g., as  Fig. \ref{fig:evolution_clusters}a). Cores that are initially bound and classified as Phase III may begin collapse and form protostars without passing through other phases (see Fig.~\ref{fig:evolution_clusters}c). In contrast, cores that are marginally bound and/or pressure confined (depending on core definition, see Appendix \ref{sec:appendix_coredef}) but not sufficiently massive to collapse likely undergo a phase of turbulent decay, developing a significant central coherent region, and evolving into Phase II. Such cores may transition between Phases I, II and III depending on their local environments and how they accrete material \citep[e.g., as described by][]{Burkert_2000, Hennebelle_2009, Hopkins_2013, Padoan_2020}.

\begin{figure*}
\includegraphics[width=1.5\columnwidth]{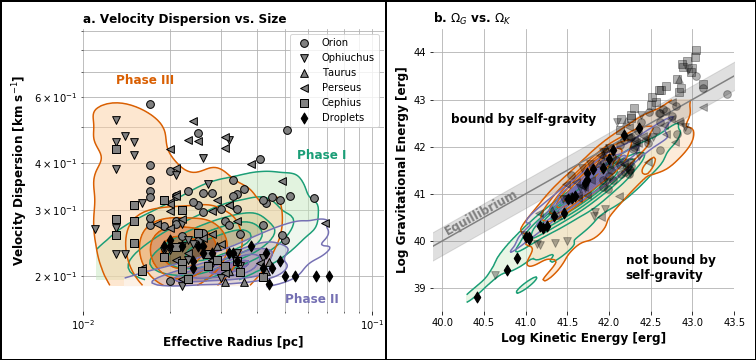}
\caption{\label{fig:compdata}  \textbf{(a)} Total velocity dispersion versus size with colors indicating their assigned phase as discussed in \S\ref{sec:results_phases}. The distribution of simulated cores in each phase is shown as contours of constant posterior probability in a Gaussian kernel density estimation (KDE) analysis that estimates the underlying probability density function in this parameter space. Cores observed in different star-forming regions are indicated by the symbols. \textbf{(b)} Same as (a) for the kinetic and gravitational potential energies.}
\end{figure*}

Due to the turbulent nature of the core environment, we find that core characteristics are non-deterministic. Cores in all three phases may disperse (Fig.~\ref{fig:phase_evolution}, see also \citealt{Smullen_2020}). This suggests that the location of an observed core in the parameter space does not uniquely determine   whether it will survive or become protostellar. Cores with significant coherent regions are more likely to live longer but are also not guaranteed to form stars at a later time ( e.g., Table \ref{table:classification_cores}). This suggests that many observed starless cores may not in fact go on to form stars. For example, our results suggest that low-mass cores with initially high virial ratios, such as a subset of Orion and Ophiuchus cores that appear towards the top of Phase I (see Fig.\ref{fig:compdata}) have a high likelihood of dissipation within $\sim 2\times10^5$ years. 

The exact percentages for the survival rates likely depend on the degree of clustering and cloud physical conditions \citep[e.g.,][]{Guszejnov_2022}. However, the fact that some cores not bound by self-gravity continue to evolve and may eventually become prestellar/protostellar is consistent with the  substantial number of observed unbound cores.  \citet{Chen_2019a} found that (Phase II) coherent cores, not bound by self-gravity, are instead confined by turbulent motions of the ambient gas. Similarly,  Orion and Ophiuchus contain a large number of unbound cores, which can be explained by a significant confining pressure \citep{Kirk_2017a,Kerr_2019}.  This confinement, provided by the turbulent pressure of the ambient gas, helps explain why many apparently unbound cores persist and some eventually become protostellar (e.g., Fig~\ref{fig:evolution_clusters}bc).  Our analysis suggests that the degree of unboundedness may be due in part to the fiducial core definition,  which focuses on an inner compact portion of the core and misses a substantial part of the core mass (see Appendix \ref{sec:appendix_coredef}). 
However, we caution that even if confining pressure helps to explain the existence of the large number of such structures, our results imply that many of these will not go on to form stars.

Cores inhabiting Phase III  have the highest likelihood both of persisting ($30$\%) and of being protostellar ($23$\%). This suggests the subset of observed starless cores in Ophiuchus, Orion and Perseus mapped to Phase III prototypes will become protostellar. Based on our tracks this may occur within $\sim 1-2 \times 10^5$ years, although the timescale for the evolution is difficult to constrain from the placement within the UMAP alone.

Overall, cores appear to transition smoothly between phases  as evinced by the significant amount of time cores often spend in one prototype and one phase before moving to another (e.g., Fig~\ref{fig:tsne_timescale}) and the concentration of tracks in limited parts of the parameter space (e.g., Fig~\ref{fig:evolution_clusters}).  As discussed above, the appearance and growth of coherent regions appears to be gradual, and a core remains not bound by self-gravity in parts of Phase II.  On the other hand, the transition between Phase II and Phase III  or Phase I and Phase III corresponds to a shrinking or complete disappearance of the central coherent region (Fig.~\ref{fig:tSNE_var_projections}).  However, we note that there is a certain degree of overlap and that some of the Phase III cores still contain coherent regions (Fig.~\ref{fig:tSNE_var_projections} and Fig.\ \ref{fig:compdata}).  
An observational example is the star-forming coherent core in the B5 region in Perseus identified by \citet{Pineda_2010}.  This coherent core is associated with a known protostar and contains at least three other starless substructures \citep{Pineda_2015}.  \citet{Pineda_2010} observed an increase in velocity dispersion near the protostar in B5, which is also exhibited in some of the star-forming Phase III cores (Fig.\ \ref{fig:clusterprops}). This elevated dispersion could either be due to gravitational infall or the protostellar outflow.  One of the starless substructures, B5-Condensation1, also exhibits a larger central line width at higher resolution, which is likely due to infall \citep{Schmiedeke_2021}.  

Gravitational boundedness is often used to distinguish between conventionally identified {\it starless cores,}  i.e., those with no protostar which are considered unlikely to form stars, and prestellar cores, which likewise contain no protostar but are expected to become star-forming.  As shown in Fig.\ \ref{fig:tSNE_var_projections}f, there is no sharp boundary between gravitationally bound and unbound cores.  There are Phase II cores that are gravitationally bound according to the virial analysis, and there are Phase III cores that are not gravitationally bound.  Both the disappearance of the coherent region and the emergence of gravitational boundedness are related to the onset of gravitational infall in our evolutionary picture.  {\it In this dynamic picture, one should not rely on a conventional virial analysis to predict whether a core will eventually form stars or not.}

\begin{figure}
\begin{center}
\includegraphics[width=0.9\columnwidth]{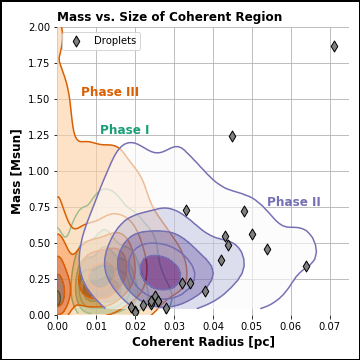}
\end{center}
\caption{\label{fig:coherent}  Core mass versus size of the coherent regions. Contours show lines of constant probability from a KDE analysis for each of the phases. Diamonds indicate droplet properties, where the droplet size is the size of the coherent region by definition. Most droplets appear to be Phase II members.}
\end{figure}

\subsection{Comparison with Low-mass Star and Core Formation Models}
\label{sec:discussion_formation}
 To date, a great deal of theoretical work has been directed towards the question:  how does a core form in a molecular cloud, and how does core formation lead to the formation of stars?  In this section we discuss three representative models of low-mass core and star formation and compare our results with these models.

Starting with \citet{Padoan_1997}, a series of works have proposed turbulent fragmentation as the dominant mechanism in forming cores \citep[see][and references therein]{LeeOffner2020}.
In this model, structures develop in a top-down sense.  Structures at smaller scales form when turbulent fluctuations in the parent larger-scale structure cause sub-regions to become gravitationally unstable.  \citet{Hopkins_2013} suggests that the physical properties of cores formed via this mechanism are set at the time of fragmentation and are only weakly modified by the collapse process.  In the \citet{Hennebelle_2008} model the decay of turbulence does not affect the selection process, which adopts gravitational instability as the criterion to select structures that continue evolving and eventually become prestellar/protostellar.  In contrast, we find that the evolution of turbulence within the core plays an important role.  As discussed in \S\ref{sec:discussion_evolution}, turbulence dissipation in the first $\sim$1-2$\times$10$^5$ years is necessary to reduce turbulent support before gravitational collapse starts.  Although we do find that some Phase I cores are close to being gravitationally unstable and evolve directly into Phase III, we find that focusing only on density structures that are above the collapse threshold would bias the analysis by excluding cores that eventually become star-forming.  However, based on our analysis, we agree that turbulent fluctuations are important in creating the initial distribution of density structures, although unlike in the theoretical framework of turbulent fragmentation, these density structures do not need to be initially gravitationally unstable to continue evolving to become prestellar cores.

\citet{CChen_2014}, \citet{CChen_2015} and \citet{CChen_2018} examine the formation of cores in the post-shock layers of supersonic converging flows.  In their model, the converging flows collide in a plane-parallel fashion.  \citet{CChen_2015} find that cores and filaments form simultaneously in these post-shock layers.  The cores have subsonic velocity fields not unlike the Phase II coherent cores, as a result of the assumption that the turbulence has already been dispersed on small scales due to the initial conditions \citep[e.g., see Fig.\ 5 in][]{CChen_2016}.  They find that although the subsonic cores are initially not bound by self-gravity, anisotropic flows \citep[referred to as anisotropic contraction in][]{CChen_2014} along directions parallel to the post-shock layers help the subsonic cores collect mass.  The anisotropic flows continue to add mass to the cores, even after the cores become gravitationally unstable and collapse starts.  Generally speaking, the process examined by \citet{CChen_2015} corresponds to the evolution of a subset of our Phase II cores toward Phase III.  They find that the timescale of the anisotropic phase, which starts when the anisotropic flows emerge and ends when the cores become gravitationally unstable, is 2$\times$10$^5$ to 3$\times$10$^5$ years, comparable to our Phase I/II + Phase III mean lifetimes.  
These works by \citet{CChen_2015} demonstrate that converging flows can be an efficient way to dissipate turbulence, although in reality, the idealized setup of cloud-scale plane-parallel converging flows is unlikely in turbulent clouds.  A similar process involving converging flows may explain the formation of the dense filaments and the cores within them that  we also observe here.  However, their setup alone cannot fully explain the formation and evolution of isolated Phase I and Phase II cores outside the filaments, which appear to be correlated with mild and local shock-induced features in our model (see Fig.~\ref{fig:dendro_phase}). These isolated cores collect mass as they move across the turbulent cloud without need for converging flows.  Future studies of cloud-scale converging flows in more realistic settings within turbulent clouds are needed to understand their effects on core evolution and turbulence dissipation.

 \citet{Semadeni_2017} and \citet{BallesterosParedes_2018} propose a gravity-regulated model of core formation, where dense cores form via hierarchical gravitational fragmentation.  In the analytical model put forward by \citet{BallesterosParedes_2018}, a star-forming core starts its evolution in a state of gravitational instability and remains gravitationally unstable throughout the evolution. Thus, a core in this model undergoes gravitational collapse at all times.   \citet{BallesterosParedes_2018} propose that outside-in gravitational collapse generates the distribution of velocity dispersions observed in coherent cores, with larger velocity dispersions at larger radii and smaller velocity dispersions in the core centers. The simulated core in this model develops a density profiles similar to the critical Bonnor-Ebert sphere, with $\rho$ $\propto$ $r^{-2}$.  Based on our analysis, we conclude this model lacks the ability to explain the turbulence in Phase I cores and the dissipation of turbulence during Phase I and Phase II.  In our analysis, when a core evolves from Phase II to Phase III, gravitational collapse starts at the center of the core \citep[an inside-out collapse as proposed by][]{Shu_1977}, raising the velocity dispersion at the center above the thermal sonic speed first before increasing the gas dispersion towards the core edges.  This can be seen in Fig.\ \ref{fig:clusterprops}, where many of the Phase III cores have centrally enhanced velocity dispersions.  As discussed above, most Phase I cores and Phase II cores have density profiles that are shallower than a critical Bonnor-Ebert sphere, although at later times, the profiles do approach Bonnor-Ebert-like profiles with $\rho$ $\propto$ $r^{-2}$.  On the other hand, \citet{Semadeni_2017} show that hierarchical gravitational fragmentation is capable of creating star-forming cores that have physical properties similar to those of the observed cores in a study of core formation in a molecular cloud undergoing global gravitational collapse in simulations.  However, similar to the analytical model presented by \citet{BallesterosParedes_2018}, the cores in the simulations studied by \citet{Semadeni_2017} appear to be gravitationally supercritical at all times, while in our model, most cores form as subcritical structures,  whose evolution is driven by the details of their formation from the turbulent cloud environment.  The gravity-regulated model cannot fully explain the evolution of cores seen in our analysis.

In summary, the underlying difference between the picture presented in this paper and previous theoretical models is the inclusion of gravitationally subcritical structures in the core evolution theory.  In previous models, subcritical density structures are excluded in the analysis under the conventional assumption that such structures disperse before they can become prestellar/protostellar.  Our model shows otherwise.  As discussed in \S\ref{sec:results_evolution}, we find that a portion of cores that are not bound by self-gravity continue to evolve and eventually become prestellar/protostellar.  Critically, turbulence dissipation appears to constitute an important separate stage of core evolution.
Future studies that examine gravitationally subcritical cores along with supercritical ones are needed to understand the process of turbulence dissipation and how it sets the initial conditions for the later phase of gravitational collapse and star formation.

\subsection{Comparison with High-mass Star Formation Models}
\label{sec:discussion_formationhigh}
 Our simulation represents typical nearby low-mass star-forming regions, like Perseus, Ophiuchus and Taurus, with similar gas temperatures, column densities and velocity dispersions.  Likewise, the simulated core properties, including masses and sizes, are similar to those of cores identified in these regions.  This reinforces that our proposed core evolution model is applicable in the context of low-mass star formation as defined by stars with masses below a few solar masses.  High-mass star formation, which is characterized by higher gas temperatures, velocity dispersions, column densities and stellar densities, may proceed very differently and not pass through the phases we propose here.  However, observations suggest star formation exists on a continuum, low and high-mass star formation occurs co-spatially and contemporaneously, and there is not necessarily a clear dichotomy between them.  To date, no coherent cores with high masses that could be progenitors of massive stars have been observed.  This may be because such cores are distant and rare or because few, if any, massive starless cores exist \citep{Tan_2014}.
However, our evolutionary model shares some characteristics with several models for high-mass star formation, as we discuss here.  During Phase I cores are trans-to-supersonically turbulent and appear to be supported by turbulent pressure, characteristics that are adopted as the initial conditions of massive cores in the Turbulent Core (TC) model for high-mass star formation \citep{McKee_2002, McKee_2003}.  In this model, turbulence provides internal pressure support and mediates gravitational collapse.  Later work notes that strong magnetic fields may also contribute to the stability of massive cores \citep{Tan_2013}.  However, the TC model does not address in detail how such cores form.  The challenge of identifying truly massive, starless cores and the apparent rarity of such objects suggest that some degree of collapse and star formation proceeds before a large reservoir of gas accumulates \citep{Padoan_2020,Grudic_2022}.  In other words, massive star formation is contemporaneous with massive core formation.  In our model a significant portion of the core mass accumulates before the internal turbulence decays and collapse proceeds.  However,  the mass becomes more centrally concentrated during Phase III, suggesting that some degree of core growth continues during the collapse phase  but may not be included within the FWHM boundary (see Appendix \ref{sec:appendix_coredef}).

In the opposite extreme, the competitive accretion (CA) model predicts that cores as discrete objects are relatively unimportant to the final outcome of star formation \citep{Zinnecker_1982, Bonnell_2001a, Bonnell_2001b}.  Instead, massive stars form at the center of clouds within the largest gravitational potential well, which funnels material inwards and facilitates high stellar accretion rates.  In this case, core masses are independent of the final masses of the stars that form within them, and massive starless cores never exist \citep{Smith_2009, Mairs_2014}.  The CA model stresses the importance of the local environment and role of neighboring stars. In our model, cores form both outside and inside filamentary regions, where the latter has the greatest ability for cores (and protostars) to grow due to inflowing gas.  We find that Phase III cores tend to have closer near-neighbors, $\bar d= 0.13^{+0.06}_{-0.05}$ versus $\bar d= 0.17^{+0.1}_{-0.07}$ and  $\bar d=0.18^{+0.13}_{-0.07}$ (see Table \ref{table:properties}) for Phase I and II cores. This suggests that environment has some influence on the progression of core evolution.  The difference in clustering between Phase I/II cores and Phase III cores may be in part because some fraction of cores disperse before reaching Phase III, which could be more likely to occur if the local environment does not allow sufficient mass accretion to trigger collapse. 

Recently, \citet{Padoan_2020} proposed the inertial-inflow model, in which massive stars form in turbulent regions characterized by large-scale converging flows.  The inertial-inflow model is formulated by analyzing magnetized, driven turbulent simulations not too dissimilar from the one we analyze here, although \citet{Padoan_2020} follow a larger spatial volume and do not resolve the formation of low-mass stars (M$_*$ $\lesssim$ 2 M$_\odot$).  Turbulent fragmentation produces the initial core properties and sets their growth timescale; massive stars form in cores that continue to grow through accretion.  This model predicts that truly massive starless cores do not exist, since collapse begins before a significant amount of mass accumulates.  Similarly, \citet{Grudic_2022} find a very dynamic picture for high-mass star formation, in which massive stars require a long time ($\gtrsim 1$\,Myr) to reach their high masses and these stars accrete at increasingly high rates. Of the high-mass models we discuss here, these two models are the most similar to the one we propose for low-mass star formation, namely, in that it emphasizes the dynamic nature of core evolution. However, it does not explicitly address the early stages of core formation, and the cores identified in the simulation are gravitationally bound by construction, so they are most analogous to our Phase III cores.  It seems possible that  turbulent decay and the formation of coherent regions play an important role in low-mass star formation as we propose here (e.g., Figure \ref{fig:coherent}),  and the inertial-inflow model represents a natural extension of core evolution for higher mass stars.  Future work is required to determine how the Phases we identify here relate to high-mass core formation and evolution.

\subsection{Observational Identification of Core Phases}
\label{sec:discussion_observation}

 Intriguingly, coherent cores have only been directly observed and resolved using observations of NH$_3$ hyperfine line emission.  Meanwhile, there are observations of C$^{18}$O and N$_2$H$^+$ molecular line emission that either did not resolve the transition to coherence and/or probed only the interior of a coherent core \citep{Goodman_1998, Caselli_2002}. Our models suggest that many starless cores contain compact coherent regions that are below the current observational resolution.  By comparing the profiles in  Fig.\ \ref{fig:clusterprops}, we see that the transition to coherence generally corresponds to a density threshold of $\geq$ 2$\times$10$^4$ cm$^{-3}$ and that most such cores have peak densities below 10$^5$ cm$^{-3}$, which may make them difficult to detect. In addition, extended coherent regions may be hidden in observations due to the embedding turbulent gas \citep{Choudhury_2021}.

Phase I cores have similarly low peak densities and properties; without sufficiently high resolution (e.g., $\lesssim 0.01$\,pc) it would be observationally difficult to distinguish between Phase I and Phase II cores.  Molecular line tracers that are also sensitive to lower densities would make the observed line widths appear broader due to the turbulent motions of the lower-density materials along the line of sight. Consequently, it would be difficult to identify and resolve an internal coherent region.  Molecular line tracers tracing higher densities would resolve the interior of the coherent region but not the transition to coherence occurring at $\geq$ 2$\times$10$^4$ cm$^{-3}$ at the same time \citep[this may be the case for the N$_2$H$^+$ observations performed by][]{Caselli_2002}. 

In contrast, Phase III cores are relatively easier to detect.  They are expected to be denser and more chemically evolved, providing a larger selection of possible molecular line tracers.  These properties likely account for the larger number of observed gravitationally bound prestellar and protostellar cores compared to coherent cores.  Probing the internal velocity structures of Phase III cores is usually limited by the saturation threshold, and choosing the right molecular line tracer becomes critical.  Numerous examples of prestellar and protostellar cores that likely correspond to this phase in the simulations have been identified in observations \citep{Tafalla_2004, Enoch_2008, Kauffmann_2008, Rosolowsky_2008a, Belloche_2011}.  At an even later stage, the formation of protostars within cores provides an extra observational hint that they belong to Phase III such as excess infrared emission and/or molecular outflows \citep{Bontemps_1996, Arce_2007}.

The starting time of a core is subject to uncertainty in the definition of a core.  In our analysis, cores are defined by the parameters of the dendrogram identification algorithm  and FWHM criterion, and we expect that choosing slightly different parameters would yield slightly different core properties.  As described in \S\ref{sec:analysis_id}, we require a density structure to have a size larger than $\sim$0.028 pc above a density threshold of 10$^4$ cm$^{-3}$ to be identified as a core.  In reality, the growth of a density structure in the molecular cloud starts before gas reaches these densities.  The growth time before we identify the core may be estimated with the free-fall time, $t_\mathrm{ff} = \sqrt{3\pi / 32 G\rho}$, which is 3.1$\times$10$^5$ yr for a density of 10$^4$ cm$^{-3}$.  Processes such as the formation of complex molecular species likely start during the initial growth of the density structures and before the core is classified into one of the three Phases we define here, but the formation time of different species varies and abundances may not reach a detectable level until the core remains above 10$^4$ cm$^{-3}$ for $\sim 10^5$ yr \citep{Suzuki_1992,GAS_DR1}.

\subsection{Comparison Caveats}
\label{sec:caveats}

 In this section we discuss several caveats to our analysis and comparison to observations.

First, our simulation does not include stellar feedback. Feedback, particularly in the form of protostellar outflows, appears to be critical in setting both the local core-to-star and global cloud-to-star efficiencies \citep{Federrath_2015,Offner_2017,Grudic_2022}. Feedback is also responsible for driving turbulence over a range of scales within molecular clouds \citep[e.g.,][]{Offner_2014,Offner_2018}. The star-forming regions we compare with in this work appear to have ubiquitous feedback in the form of outflows and winds \citep[e.g.,][]{Xu_2020a,Xu_2020b,XuOffner_2022}. Consequently, we expect the presence of feedback to alter the simulation core properties and their cloud environment to some degree. In comparing with observations, we mitigate the lack of feedback in the simulation in two main ways. First, we compare to NH$_3$ observations, which trace denser gas, where the imprint of feedback is small. Protostellar cores observed with dense-gas tracers have relatively low (sub- or trans-sonic) velocity dispersions \citep{Kirk_2007,Rosolowsky_2008a}.  The signature of feedback in NH$_3$ line widths  at higher resolution is also usually small as in the case of B5, which hosts a Class I protostar \citep{Pineda_2015}.
Second, the large majority of the observed cores that we compare with are thought to be starless. Thus, while  stellar feedback will likely alter the details of the prototype learning and UMAP visualization, we expect it will have little effect on the resulting classification and our general conclusions.

Protostellar outflows also regulate core lifetimes by entraining and expelling dense material. Simulations with feedback find that the lifetime of protostellar cores, as defined by when most accretion occurs, is $\sim 2 \times 10^5$ yr \citep{Offner_2017}, albeit with a large amount of scatter \citep{Grudic_2022}. Only one of the protostellar cores in the simulation disperses by the end of the calculation (from Phase III). Without feedback the protostellar core lifetime and more generally the time star-forming cores spend in Phase III (
5.0$\pm$0.4$\times$10$^5$ yr, see \S\ref{sec:results_evolution}) is over-estimated, since there is no mechanism to halt additional gas accretion onto a core and protostar.

We also caution that the simulation models core evolution under one set of initial conditions. These conditions represent the gas temperatures, densities and velocity dispersions typical of conditions in nearby low-mass star-forming clouds. Although we find these conditions produce cores with properties in good agreement with those of observations (e.g., Fig.~\ref{fig:compdata} and \ref{fig:coherent}), further work is required to determine the impact of variations in mean magnetic field, density, velocity dispersion and cloud geometry on core formation and evolution \citep[e.g.,][]{Guszejnov_2021,Guszejnov_2022}.

In addition, we do not carry out synthetic observations of the simulations, which are required for true "apples to apples" comparisons between models and observations \citep{Haworth2018,Rosen2020}. This would require calculating the NH$_3$ abundances using chemical networks or adopting an abundance model \citep[e.g.,][]{Offner_2013,Gaches_2015,GAS_DR1}, performing radiative transfer calculations to model the emission \citep[e.g.,][]{Beaumont_2013,Gaches_2015} and accounting for observational resolution \citep[e.g.,][]{Bradshaw2015,Sokol_2021}.  We mitigate the impact of these uncertainties by focusing on cores observed in NH$_3$, which has a low volume filling factor within local clouds and thus suffers less from projection effects that otherwise produce chance alignments of over-densities along the line-of-sight. We also calculate the properties of the simulated cores using a grid resolution comparable to the GAS pixel resolution of the observed star-forming regions. Despite this, our approach does not fully encapsulate the uncertainties in the observational data. Future work analyzing the evolution of cores in the space of synthetic NH$_3$ observations is required to more securely map the observations to the simulated data.

Finally, as discussed in \S\ref{sec:obs_comparison}, we project the observations into the simulation space using a subset of the core properties. A more complete comparison requires including the radial profiles of the observed cores in the prototype matching. However, these data have not been derived for cores in most of the catalogs we compare with. This additional information would help disentangle high velocity dispersions produced by infall motions from those produced by core turbulence. Our prototype learning makes this distinction easily, cleanly separating protostellar cores, which are experiencing infall (Phase III), from cores that are simply very turbulent (Phase I; see Figure \ref{fig:clusterprops}). However, the set of observed bulk core properties may be insufficient to identify this distinction. For example, in Figure \ref{fig:observed_cores} a number of cores in Ophiuchus, Perseus and Orion are mapped into the upper part of Phase III, where the simulated protostellar cores reside. Most of these observed cores are not (currently) associated with any identified infrared source, so we cannot determine whether their placement there indicates incipient star-formation or whether it indicates only that they have a high degree of turbulence. The latter scenario would suggest some of these are more analogous to our Phase I cores, which are less likely to become star-forming. Future catalogs of core properties that include velocity dispersion and column density profiles will enable methods like this one to better distinguish between these two possibilities.

\section{Conclusions}
\label{sec:conclusions}

We present a method to identify, track and characterize the evolution of dynamic gas structures in simulations. Our method is general and is applicable to other numerical models of star formation. Unlike many previous core identification and analysis methods, we do not make a priori assumptions about the physical properties of the cores or their density and velocity dispersion distributions.

To provide a complete picture of core formation and evolution that links turbulent molecular clouds to star-forming cores, we study the formation, evolution and collapse of dense cores identified in an MHD simulation.  We identify all independent density structures above 10$^4$ cm$^{-3}$ in the simulation using the dendrogram algorithm. For each core we construct a data vector comprised of the density and velocity dispersion profiles,  core mass, radius, coherent region radius, total velocity dispersion, density exponent, kinetic energy and gravitational energy. We utilize 
prototype learning to characterize the core data features, FCM to cluster the data, and UMAP to project the information into a two-dimensional space.  We then track the cores as they evolve and move across both the simulation and the learned prototype space.  As a result, we find
three distinct evolutionary phases.  Phase I 
 represents unbound turbulent structures; we refer to this phase as the turbulent phase. Since these cores are unbound, they must gain mass or become quiescent in order to form stars. Phase I cores have turbulent internal velocity dispersions and shallow density profiles.  Phase II corresponds to the dissipation of turbulence and the formation of an extended coherent region, which is defined as a region with subsonic and nearly uniform velocity dispersion.  Phase II cores resemble observed coherent cores, including ones that are not bound by self-gravity like the droplets observed by \citet{Chen_2019a}.  We refer to this phase as the coherent phase.  Phase III cores are characterized by gravitational infall, which often dominates the internal dynamics.  Phase III cores include both gravitationally bound prestellar and protostellar cores.  They also tend to be more compact and lie in more clustered regions.  About 23\% of these cores contain protostars, such that this group contains 96\% of the protostellar cores. Consequently, we refer to Phase III as the prestellar/protostellar phase. We estimate typical lifetimes of

1.0$\pm$0.1$\times$10$^5$ yr, 
1.3$\pm$0.2$\times$10$^5$ yr, and 
1.8$\pm$0.3$\times$10$^5$ yr,

respectively, for Phase I, II and III.  

We track the evolution of cores through prototype space and examine how they evolve through the Phases over time. Overall, we find that core evolution is dynamic with 
85$\pm$4\% 
of cores changing phase at least once or dispersing during their lifetimes.  In addition, the instantaneous properties of a given core are not predictive of its eventual evolution; cores do not follow one single evolutionary path through the three identified phases. We attribute this to a combination of truly stochastic processes, such as ongoing gas accretion and interactions with the turbulent cloud environment as well as with other cores, and ambiguity about the core boundary location, which does not always capture all the associated gas. Of the cores we identify and track, 37\% disperse before becoming self-gravitating and 32\% merge with another core. This suggests that most observed starless cores have highly uncertain futures and many will not go on to form stars. 

However, we are able to identify some general trends for different core populations. We find that cores that are {\it short-lived} and exist for only two snapshots before dispersing  primarily belong to Phase I or II. The subset of {\it long-lived} cores that exist for all snapshots appear to cycle through adjacent regions of Phase I, II and III space, spending a significant fraction of their lives as quiescent Phase II coherent cores. Finally, cores that form protostars can begin in any of the three phases but spend most of their lives in Phase III, where they remain once they become protostellar.  As prestellar cores these structures evolve upwards and to the right in the UMAP space, until they reach the region of Phase III parameter space where nearly all protostellar cores reside.

 We compare our simulated cores to observed cores detected in NH$_3$ emission in the Taurus, Cepheus, Orion, Perseus and Ophiuchus star-forming regions by the Green Bank Ammonia Survey  \citep[GAS][]{GAS_DR1,Kirk_2017a,Kerr_2019,Keown_2019,Chen_2019a}. After excluding cores with gas temperatures $\geq 15$\,K, we demonstrate that the simulated and observed cores have similar core masses, sizes, velocity dispersions and virial ratios. We map the observed cores into the prototype space and project them onto the two-dimensional UMAP visualization derived from the simulated cores. We show the observed cores are matched to core prototypes in all three phases. We estimate that at least 20\% of these will disperse, while $\sim 50$\% will go on to form stars. The remaining 30\% map to long-lived quasi-equilibrium structures whose final evolution is ambiguous. 

We find that the coherent cores observed by \citet{Chen_2019a} are primarily classified as Phase II.  The core evolution paths we identify indicate that coherent cores represent an important, earlier stage of evolution for many prestellar and protostellar (Phase III) cores.  We demonstrate that the observations of NH$_3$ hyperfine line emission with a physical resolution of $\sim$0.2 pc or finer, like the ones carried out by \citet{GAS_DR1}, are ideal for detecting Phase II cores.  However, the simulations suggest that many observed cores mapped to Phase I and some in Phase III likely host a compact coherent region, $R_{\rm coh}\lesssim 0.02$\,pc, that remains unresolved. We find a number of cores in  Taurus, which is a relatively quiescent region, are also classified as Phase II cores. Follow-up examination of the velocity profiles of these cores may find evidence of a coherent sub-region. In contrast, cores detected in Orion, Perseus (specifically in NGC 1333), and Ophiuchus have higher velocity dispersions and are predominantly classified as Phase I or III. 

 Future work is needed that examines simulations with more diverse initial conditions and additional physics to evaluate the impact of cloud properties and stellar feedback on core evolution. 
 
\section*{Acknowledgements}
This work was supported by Cottrell Scholar Award \#24400 from the Research Corporation for Science Advancement, NSF CAREER 1748571 and NSF AAG 1812747 and 2107942.  AG acknowledges support from the NSF via AST 2008101 and CAREER 2142300.
JEP acknowledges the support by the 
Max Planck Society. 
The authors thank anonymous referees and Mordecai-Mark Mac Low for comments that significantly improved the manuscript and acknowledge helpful discussions with Michelle Ntampaka and Keith Hawkins.   
This research made use of Astropy, a community-developed core Python package for Astronomy \citep{astropy}.

\section*{Data Availability}

The data supporting the analysis and plots in this article are available by request to the corresponding author. A public version of the {\sc orion2} code is available at \url{https://bitbucket.org/orionmhdteam/orion2_release1/src/master/}.
 
\bibliographystyle{mnras}
\bibliography{main.bib, ml.bib}

\appendix

\section{Down-Selecting the Core Property Vector}
\label{sec:appendix_lasso}

\begin{figure}
\includegraphics[width=0.95\columnwidth]{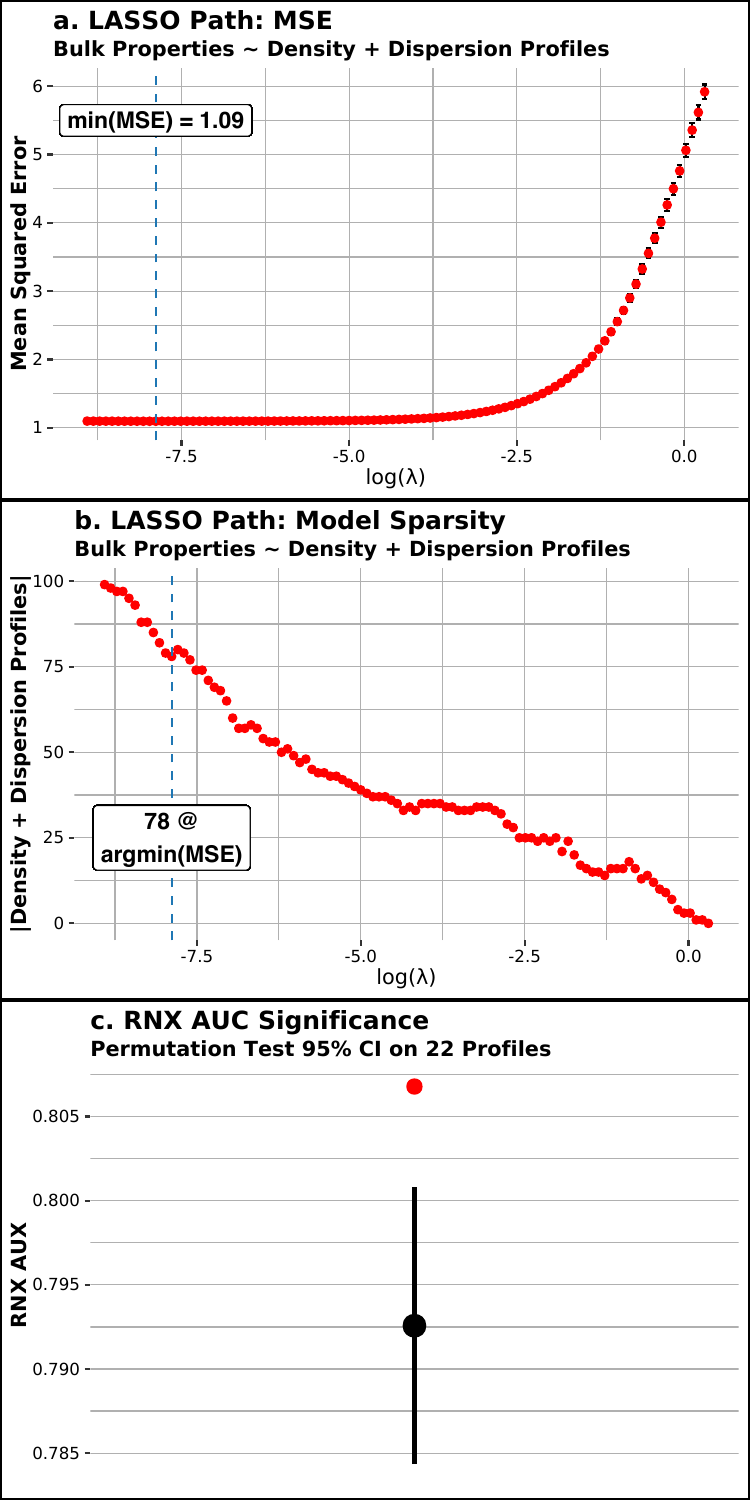}
\caption{\textbf{(a)}  LASSO Mean Squared Error (as a function of penalty parameter $\lambda$) of regressing bulk properties $\sim$ density + dispersion profiles. Minimum MSE occurs at $\log(\lambda) \approx -7.8$, which corresponds to keeping 78 profiles in the regression model that are most associated with the bulk properties, as shown in \textbf{(b)}. The complement of this set (22/100 profiles) contains information with less association with the bulk properties, which we include in our analysis. \textbf{(c)} The RNX AUC topology-preservation measure (red point, from \citealt{Lee2015MultiscaleSI}) of the LASSO-selected set of 22 profiles, compared to a confidence interval for RNX AUC (black point range) obtained via a permutation test of selecting 22 profiles at random. LASSO RNX is statistically larger than chance selection, indicating our selection procedure has kept information relevant for inferring structure from the dimensionally-reduced dataset.  }\label{fig:lasso}
\end{figure}

To remove duplicate information from the core profiles we use the Least Absolute Shrinkage \& Selection Operator \cite[LASSO,][]{Tibshirani1996}, which employs penalized linear regression to fit a model $Y \sim X\beta$ according to the following:
\begin{eqnarray}
\min_{\beta} \, SSE &=& ||Y - X \beta||_F \\
{\rm subject~to~} ||\beta||_1 &\leq& \lambda,
\end{eqnarray} 
where SSE is the Sum of Squared Errors of the regression, and $||\cdot||_F$ is the Frobenius norm. In our case, $X$ is the 100-d concatenated density and dispersion profiles while $Y$ contains the 6 bulk properties for each core, meaning our $\beta_i$ are 6-dimensional vectors of regression coefficients. 

For a given value of $\lambda$ optimization of the above forces some set of $\{\beta_i\} \to 0$, indicating removal of variable(s) $\{i\}$ has minimal impact on model SSE. The optimal value of $\lambda$ is typically selected from a grid of candidate values via cross-validation, using SSE or Mean Squared Error as a guide. From Figure \ref{fig:lasso}[a], our model MSE is minimized at $\log(\lambda) \approx -7.8$, which corresponds to retaining 78 of the 100 concatenated profiles in the model (model sparsity as a function of $\lambda$ is given in Figure \ref{fig:lasso}[b]). 

The lack of a sharp minimum in the MSE curve in Figure \ref{fig:lasso}[a] indicates the LASSO regression is relatively stable over a wide  range of $\lambda$ or, equivalently, that only a small subset of the concatenated density + dispersion profiles possess significant linear predictive power for the bulk properties, collectively.
Because we include the bulk properties in our analysis, we are more interested in the \textit{complement} of the set of LASSO-selected profiles, i.e., variables that contain information \textit{other} than what can be found in the bulk properties. Thus, we retain only 12 density + 10 dispersion = 22 profile variables for analysis. We note that a more sparse LASSO model, which Figure \ref{fig:lasso}[a] suggests is statistically equivalent, has a larger set complement. As our goal with this preliminary analysis is to reduce the number of profiles used for prototype learning and clustering, we have chosen the optimal model sparsity corresponding to the recommended $\argmin MSE$ to produce the largest impact for subsequent analysis.

Literature on Dimensionality Reduction (DR) algorithms often evaluate their performance via measures of topology preservation, which report how well local data neighborhoods in high-dimensional space are preserved when represented in a lower-dimensional space. As our LASSO-based variable selection is essentially a statistical DR technique, we borrow a measure known as Rescaled Neighborhood Area Under the Curve (RNX AUC, detailed in \cite{Lee2015MultiscaleSI}, equation 17) to assess the impact of removing parts of the density and dispersion profiles.  RNX AUC reports a chance-corrected proportion of $K$-nearest high(100)-dimensional neighborhoods that are preserved in low(22)-dimensional, averaged over all possible values of $K$. A value = 1 indicates the ideal case, where all neighborhood relationships are preserved at all scales in a low-dimensional representation, and values $<$ 1 signal neighborhood misalignment. Representing the 100 core profiles with the 22 selected via LASSO results in an RNX value $\approx 0.81$. To test whether 0.81 is significant, we performed a permutation test by randomly selecting 22/100 profiles, computing RNX, and repeating 100 times. A 95\% confidence interval from this non-parametric test is shown in Figure \ref{fig:lasso}[c] (black point range), alongside the LASSO-selected RNX value (red point). There is a small but statistically significant improvement in RNX when using the LASSO-selected profile subset. 

\section{Selecting the Number of Clusters}
\label{sec:appendix_cluster}

\begin{figure}
\includegraphics[width=0.95\columnwidth]{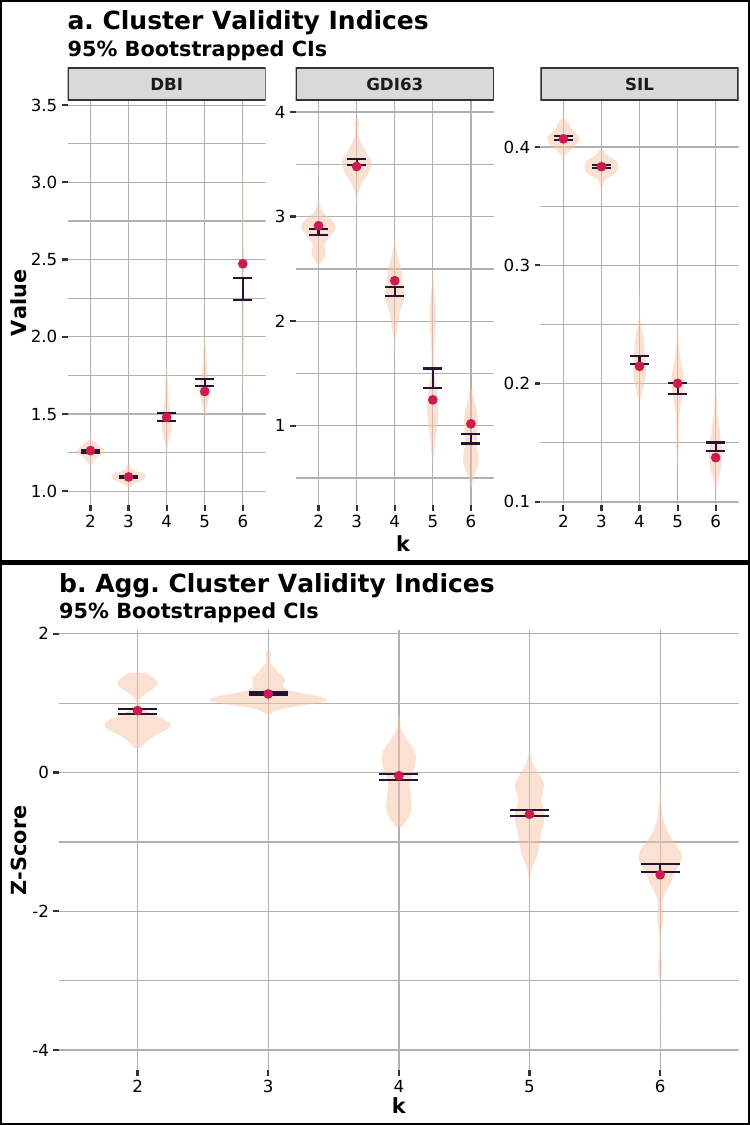}
\caption{ Bootstrapped sampling distributions (violin plots) and mean confidence intervals of each Cluster Validity Index computed as the number of $c$-means clusters $k$ ranges from 2 through 6. Red points indicate the CVI evaluated on our original (non-Bootstrapped) sample. $k=3$ attained the highest aggregate score among all CVIs (panel [b]), and the highest among two of the three measures considered here (panel [a]). Since lower values of DBI are preferable, DBI is negated before its Z-Score transformation to make it comparable to GDI63 and SIL.} \label{fig:CVI_ZScore}
\end{figure}

The $c$-means algorithm partitions data into $k$ clusters ($k$ is a user-specified parameter) regardless of whether $k$ well-defined clusters are actually present in the data. Thus, the success of $c$-means depends upon proper specification of $k$. As there is no universally superior method for determining the most appropriate value of $k$ a number of Cluster Validity Indices (CVIs) reporting the degrees of compactness and separation of clusters in a partitioning have been developed \citep{arbelaitz2013extensive}. Typically an analyst selects $k$ as the $\argmax$ (or $\argmin$, as appropriate) of a CVI computed for each clustering resulting from a range of $k$.  The weakness of such an approach is that there is, again, no universally superior CVI (the problem of choosing $k$ has been replaced with that of choosing the correct CVI for the data at hand). Consultation of several CVIs computed for a range of $k$ is an intuitive way to make this process more robust to (potentially) user-biased CVI selection, but the range and optimality conditions of each CVI vary which prohibits direct and simultaneous comparisons. Additionally, some CVIs possess an inherent bias toward a small or large $k$, e.g., the average within-cluster variance is a monotonically decreasing function of $k$ for $c$-means. 

Recent work by \citet{akhanli2020comparing} proposes a method based on resampling techniques to build an empirical sampling distribution $\hat F_{\iota(k)}$ of CVI $\iota(k)$. This sampling distribution represents values of a particular CVI $\iota$ which could result from clustering multiple datasets similar to the one originally observed, for a fixed value of $k$. The mean and standard deviation of $\hat F_{\iota(k)}$ are  used to create a standardized Z-Score of each resampled $\iota(k)$; repeating this process $B$ times for a collection of CVIs $I(k) = \{\iota_1(k),\iota_2(k),\ldots\}$ yields a collection of Z-Scores $\{z^b_{\iota_1(k)}, z^b_{\iota_2(k)}, \ldots \}_{b=1}^B$ which are directly comparable (i.e., have a similarly standardized scale), both amongst themselves and over a range of $k$. Further, the observed value of CVI $\iota^*(k)$ (resulting from the original clustering, before any resampling occurs) is also standardized according to $\hat F_{\iota(k)}$ and averaged to create an aggregate index $\bar \iota^*(k)$ bearing influence from all members of $I(k)$. The $\bar \iota^*(k)$ can now be compared across $k$, and the best clustering according to this aggregation is selected as $\argmax_k \bar \iota^*(k)$. 

We have applied the aggregation method of \citet{akhanli2020comparing} 
to build sampling distributions and associated Z-Scores of the observed values of three different CVIs for $c$-means clusterings of the core prototypes, with $k$ ranging from 2 through 6: 
\begin{enumerate}
    \item \textbf{SIL}houette Index \citep{Rousseeuw90,campello2006fuzzy} 
    \item Generalized Dunn Index with set distance $\delta_6$ and diameter $\Delta_3$, or \textbf{GDI63}, as defined in \citet{bezdekpal1995}
    \item \textbf{D}avies-\textbf{B}ouldin Index  \citep{davies1979cluster}.
\end{enumerate}
These CVIs are commonly used in practice. Higher values of SIL and GDI63 are preferable, while DBI is optimal at its minimum. 

The sampling distributions of (i)-(iii) and their aggregated Z-Score are shown in Figure \ref{fig:CVI_ZScore}. Because lower values of DBI are preferable, its scores were negated prior to aggregation.  The $k=3$ clustering achieved the highest aggregate Z-Score of 1.15, while $k=2$ achieved the next highest (0.88). Because a 95\% confidence interval around the difference in these means is strictly positive ([0.22, 0.32]), we have selected the $k=3$ clustering for the analysis in this work. We note for completeness that the $k=1$ case is not addressed by most CVIs; because our simulated data possesses at least two natural groupings (whether or not a core is identified as containing a stellar object), any $k=1$ considerations are not applicable here.

\section{UMAP Dimensionality Reduction}
\label{sec:appendix_umap}

UMAP is a non-linear dimensionality reduction technique \citep{lee2007nonlinear} to embed high-dimensional point clouds $X \subset \mathbb{R}^d$ in a lower-dimensional space $U \subset \mathbb{R}^{d'}$. In this work, $d=28$ (the 22 profiles identified in Appendix \ref{sec:appendix_lasso} + 6 bulk properties) and we specify $d'=2$ to facilitate visualization. The low-dimensional points $u_i$ are formed by minimizing the cross-entropy between distributions of pair-wise similarities in high- and low-dimensional space. The high-dimensional point  similarities are constructed from an exponentially decaying kernel while the low-dimensional similarities are governed by a parametric generalization of Student's t-distribution. The most influential user-specified parameter $\eta$ controls the number of nearest-neighbor similarities that UMAP's cross-entropy minimization attempts to preserve and can greatly influence the quality of the resulting embedding.  $\eta$ is typically selected via trial and error over a grid of candidate values as the parameterization whose resulting embedding looks best, and is assumed to grow with sample size. 

In lieu of an ad-hoc grid search under such subjective criteria, we appeal to a more data-driven specification for $\eta$ utilizing information about data topology gleaned from Neural Gas learning (\S \ref{sec:neural_gas}). A recall of data through any vector quantizer (not just Neural Gas) gives rise to the Connectivity (CONN) graph of its prototypes \citep{tasdemir2009exploiting}, whose weighted edges convey topological adjacencies of, and local distributions surrounding, the prototypes in high-dimensional space. Thus, the data inside a prototype's receptive field (the set of data it represents) combined with the receptive fields of CONN-adjacent prototypes yields a subset of data whose pair-wise  similarities are topologically relevant. We set $\eta$ equal to the average cardinalities of these sets for each prototype, which is 36 for these data.

\section{Sensitivity to Core Definition}\label{sec:appendix_coredef}

In this appendix we examine the effect of the choice of the core definition on the clustering and core properties. Instead of using the FWHM to set the core size as above, we define the core boundary as the radius where the density profile equals $10^4$ cm$^{-3}$,  which is a more physically motivated core definition. This is effectively the average radius for a core enclosed by an isosurface with $n=10^4$ cm$^{-3}$, which has the benefit of making the core size independent of the peak density. Given the very different core definitions applied to observational data, we view this analysis as a strong test of the robustness of our analysis approach. 

Table \ref{table:properties_appendix} summarizes the core properties.
We find that the distinguishing features of each phase are preserved: cores in Phase II are still coherent, nearly all of the protostellar cores are mapped into one phase (Phase III), and Phase I cores are more turbulent and unbound. However, we find that the cores overall, especially those with protostars, are more extended and more massive. The median radius, $0.07$\,pc, is also significantly higher than the median sizes of the observed cores, while the median mass, $2.1$\msun, is comparable to that of the cores identified by \citet{Keown_2019} (see \S\ref{cepheus}).

 As in the previous Phase assignments, cores in Phase I and II  have significant overlap in their properties  with similar masses, radii and virial ratios. However, cores belonging to Phase III, which contains 
 96\%
 of the protostellar cores, are now systematically larger, 
 $0.1$\,pc, 
 and more massive, 
 $6.3$\msun. They are now $\sim$4-6 times more massive than Phase I and II cores, such that mass becomes a key characteristic distinguishing Phase I/II and Phase III. The FHWM definition appears to significantly underestimate the mass associated with Phase III cores and thus misses the growth of prestellar and protostellar cores. Unfortunately, it is not possible to define cores in observations using a number density based criterion; this is one reason we adopt the FWHM boundary as the fiducial core definition. 

Despite the change in core definition and properties 
97\% 
of the cores are classified into the same phase as before. The largest change occurs for 
Phase I
cores, which increase in number by 
$\sim4$\%. 
Most of the cores that are reclassified swap between Phase I and II with 21 cores moving from Phase I to II and 37 moving from Phase II to I. This gives confidence that our core classifications are robust and largely insensitive to differences between core definitions.

\begin{figure*}
\includegraphics[width=1.6\columnwidth]{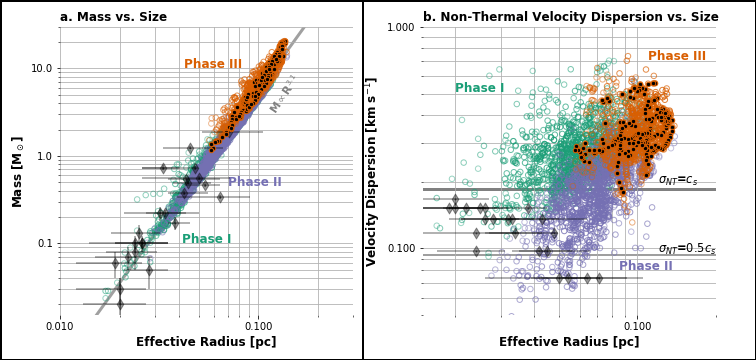}
\caption{\label{fig:properties_groups_appendix} \textbf{(a)}   Mass-size distribution of all 3,538 independent structures.  The green, purple and orange circles correspond to structures in Phase I, II and III, respectively. The symbol transparency is set by the weight of the core cluster assignment. Black filled circles indicate cores with sink particles. The grey line shows a fit to all cores. The grey diamonds represent the droplets from \citet{Chen_2019a}. \textbf{(b)} 1-d non-thermal velocity dispersion-size distribution of all 3,538 independent structures, with a color coding scheme the same as (a). The non-thermal velocity dispersion is derived for the droplets (grey diamonds) by assuming a gas temperature of 10\,K. The horizontal black lines denote the velocity dispersion values when the non-thermal velocity dispersion is equal to the sonic speed (thicker line) and half the sonic speed (thinner line) for 10~K molecular gas. Nearly all protostellar cores are members of Phase III, which tends to contain more massive and larger cores than Phase I and II. }  
\end{figure*}

\begin{figure}
\begin{center}
\includegraphics[width=0.9\columnwidth]{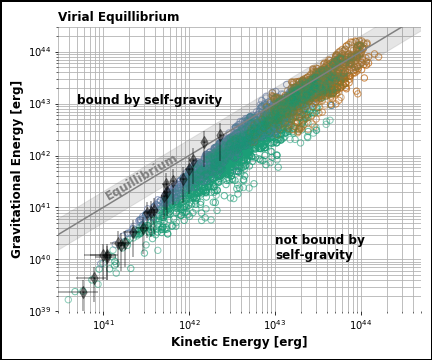}
\end{center}
\caption{ Distribution of the gravitational potential energy and the kinetic energy of all 3,538 structures where the core boundary is defined using the $n=10^{4}$cm$^{-3}$ density contour.  The green, purple and orange circles correspond to structures in Phase I, II and III, respectively.  The band from the lower left to the top right marks equilibrium between the gravitational potential energy and the internal kinetic energy (grey line) within a factor of two (grey shaded region). The droplets from \citet{Chen_2019a} are overlaid for comparison.}\label{fig:virial_groups_appendix}
\end{figure}

Figures \ref{fig:properties_groups_appendix} and \ref{fig:virial_groups_appendix} show the distributions of the core properties. In all cases, the phases show clearer separation than those identified using the FWHM definition (see the analogous Figures \ref{fig:properties_groups} and \ref{fig:virial_groups} for comparison). This suggests that a core definition encompassing more of the core envelope leads to more distinct clusters. While this core definition appears superior for clustering and classification, we instead adopt the FHWM definition in the body of the paper for the purpose of comparing more directly with the GAS data. Our analysis here suggests that the observed cores defined using {\it getsources} may miss additional material in the core envelope that would help their classification and produce more physically accurate core properties. Recovering this mass is non-trivial, since the observations are limited by the resolution, signal-to-noise and chemical characteristics of the tracers observed as discussed in \S\ref{sec:discussion_observation}. 

In Figure \ref{fig:properties_groups_appendix}a the mass-size relation is steeper with $M_c \propto R_c^{3.1}$, rather than $M_c \propto R_c^{2}$ as expected from the observed line-width size relation. In addition, the choice of boundary leads to better continuity in the properties, with the Phase III cores falling on the same, considerably tighter, mass-size relation. This suggests that underestimating the core size, or in other words adopting a core size that varies with the density peak, produces scatter in the mass-size relation. This may partially explain the very flat, high scatter mass-size relationship of the GAS data \citep[see Fig. 5 in ][for example]{Kirk_2017a}.

In Figures \ref{fig:properties_groups_appendix} and \ref{fig:virial_groups_appendix} we overlay the droplet data from \citet{Chen_2019a}, which are the core sample defined in the most similar way. The droplets are again matched predominantly with Phase II prototypes. Like Phase II cores they have small masses, sizes and velocity dispersions. While they overlap in all areas of the parameter space their sizes are systematically smaller than the median simulated core size. However, they appear to follow a similar steep mass-size relation to the simulation data.\footnote{Note that \citet{Chen_2019a} found a mass-radius power-law index of $2.4$ by combining the droplet data with updated observations of dense cores taken from \citet{Goodman_1998}, which are larger and more massive than the droplets.}  In a virial analysis, the droplets appear to follow a narrow track that hugs the distribution of simulated Phase II cores, which here are slightly offset from the Phase I distribution and closer to virial equilibrium. Nearly all of the other samples of observed cores have masses and sizes that fall outside the simulated parameter space and performing the comparison presented in \S\ref{sec:obs_comparison} is no longer a statistically rigorous or meaningful exercise.

\begin{table*}
    \setlength\tabcolsep{2.0pt} 
	\centering
	\begin{tabular}{ || l | c | c | c | c | c  | c | c |c | c | c | }
	\hline
    Core Classification & 
    $N$ &  
    $M_c$~(\msun) & 
    $R_c$~(pc) & 
    $R_{\rm coh}$~(pc) &
    p & 
    $\sigma_\mathrm{tot}$~(km s$^{-1}$) & 
    $V_{\rm bulk, 1d}$~(km s$^{-1}$) & 
    $\Omega_\mathrm{K}$/$\left|\Omega_\mathrm{G}\right|$  &
    $f_{\rm *}$\,(\%) & 
    $\bar d$\,(pc) \\ 
	\hline
	
	Phase I (Turbulent) & 
	1266 & 
	1.1$_{-0.6}^{+0.9}$ & 
	0.06$_{-0.01}^{+0.01}$ & 
	0.012$_{-0.004}^{+0.004}$ & 
	-0.87$_{-0.22}^{+0.18}$ & 
	0.33$_{-0.03}^{+0.05}$ & 
	 0.6$_{-0.2}^{+0.2}$ & 
	3.5$_{-1}^{+1.9}$ & 
	1.1 & 
0.17$_{-0.07}^{+0.10}$	 
	 \\
	
	Phase II (Coherent) & 
 1274 & 
	1.7$_{-0.8}^{+1.3}$ & 
	0.07$_{-0.01}^{+0.01}$ & 
	0.029$_{-0.006}^{+0.008}$ & 
	-0.85$_{-0.2}^{+0.15}$  & 
	 0.27$_{-0.03}^{+0.03}$ & 
	0.4$_{-0.2}^{+0.3}$ & 
	1.9$_{-0.5}^{+0.7}$ & 
	0.0 & 
	 0.18$_{-0.07}^{+0.15}$	 
	\\
	
	Phase III (Protostellar) &
	998 & 
	6.3$_{-1.9}^{+2.3}$ & 
	0.10$_{-0.01}^{+0.01}$  & 
	0.009$_{-0.009}^{+0.007}$ & 
	-1.22$_{-0.3}^{+0.22}$  & 
	0.38$_{-0.04}^{+0.06}$ & 
	0.6$_{-0.2}^{+0.2}$ & 
	1.4$_{-0.4}^{+0.5}$ & 
	22.7 & 
	0.13$_{-0.05}^{+0.06}$	 
	\\
	\hline
	
	All & 
	3538 & 
	2.1$_{-1.2}^{+2.7}$ & 
	0.07$_{-0.02}^{+0.02}$ & 
	0.016$_{-0.007}^{+0.01}$ & 
	-0.95$_{-0.25}^{+0.2}$ & 
	0.32$_{-0.04}^{+0.06}$ & 
	 0.5$_{-0.2}^{+0.3}$ & 
	2.1$_{-0.7}^{+1.1}$ & 
	6.8 & 
	0.16$_{-0.06}^{+0.1}$  
	\\
	\hline
    \end{tabular}
        \vspace{-0.1cm}
 \caption{ Physical properties of cores in each phase. We assign those that have partial membership in two different clusters to the one with the highest membership. The physical properties are measured using the density and velocity profiles derived from the dendrogram structure. The columns are number of cores and median core mass, radius, size of the coherent region, density index, total velocity dispersion, bulk velocity, ratio between the kinetic energy and the absolute value of the gravitational potential energy, fraction of members containing protostars and nearest neighbor separation. The density index is the power-law index of the function, $n = n_0 (r/r_0)^p$, fitted to the density profile of each core. The spreads are calculated using the 0.25 and 0.75 quantiles of the distribution. 
 }
 \label{table:properties_appendix}
\end{table*}

\bsp
\label{lastpage}
\end{document}